\documentclass[{epj,final}]{svjour}
\usepackage{epsf,epsfig}

\newcommand{\al}{\alpha}

\newcommand{\be}{\begin{equation}}
\newcommand{\ee}{\end{equation}}
\newcommand{\bd}{\begin{displaymath}}
\newcommand{\ed}{\end{displaymath}}
\newcommand{\si}{\sigma}

\newcommand {\fns } {\footnotesize   }

\newcommand{\bra}{\langle}
\newcommand{\ket}{\rangle}

\newcommand{\extr}{~{\rm extr}}

\newcommand{\pprime}{{\prime\prime}}
\newcommand{\ppprime}{{\prime\prime\prime}}

\newcommand{\order}{{\cal O}}
\newcommand{\minus}{\!-\!}
\newcommand{\plus}{\!+\!}

\newcommand{\bm}{\mbox{\boldmath $m$}}
\newcommand{\bn}{\mbox{\boldmath $n$}}
\newcommand{\br}{\mbox{\boldmath $r$}}

\newcommand{\bx}{\mbox{\boldmath $x$}}
\newcommand{\by}{\mbox{\boldmath $y$}}

\newcommand{\cZ}{{\cal Z}}

\newcommand{\vsp}{\vspace*{3mm}}

\begin{document}

\title{Solvable Lattice Gas Models of Random Heteropolymers at Finite
Density: II. Dynamics and Transitions to Compact States}
\titlerunning{Solvable Lattice Gas Models of Random Heteropolymers II}


\author{H.~Chakravorty,
J.~van~Mourik\thanks{\emph{Present address:}
The Neural Computing Research Group, Aston University, Birmingham B4 7ET, UK}
\and A.C.C.~Coolen}
\institute{
Department of Mathematics, King's College London, The Strand, London WC2R 2LS, UK\\
\email{hirak@mth.kcl.ac.uk}}
\date{\today}

\abstract{
In this paper we analyse both the dynamics and the high density physics of
the infinite dimensional lattice gas model for random heteropolymers
recently introduced in \cite{jort}. Restricting ourselves to site-disordered
heteropolymers, we derive exact closed deterministic evolution equations for a
suitable set of dynamic order parameters (in the thermodynamic limit),
and use these to study the dynamics of the system for different choices of the
monomer polarity parameters. We also study the equilibrium properties of the
system in the high density limit, which leads to a phase diagram exhibiting
transitions between swollen states, compact states, and regions with partial
compactification. Our results find excellent verification in numerical simulations,
and have a natural and appealing interpretation in terms of real heteropolymers.
}

\maketitle

\section{Introduction}

The practical advantage of infinite dimensional models for
random heteropolymers over the more standard ones
(which are either defined on finite dimensional lattices, or in real space)
is the absence of the polymer chain constraint.
In infinite dimensional (or `lattice gas') models, where every site is defined to be
a neighbouring site of every other site, the constituent monomers
of a polymer are given
unrestricted freedom to occupy positions. The movements of the latter are
controlled only by energetic considerations, by the maximum number of monomers allowed to
occupy any given site, and by trivial restrictions on the number of moves
allowed to occur simultaneously.
 In a recent paper \cite{jort},
which was devoted to the equilibrium solution of a particular
infinite dimensional model,
this property was shown to enable an
exact solution, in spite of the presence of
quenched disorder. (viz. the realization of the constituent
monomers of the chain, which are assumed to come in different species. 
The quenched disorder is entirely specified by the proportions of 
the various monomeric species present. No further aspects of the 
system are quenched).
In those cases where one's interest is mostly in issues which are
not directly related to spatial properties of the heteropolymer,
the compromise of infinite dimensionality might be a price worth
paying for exact solvability.

The present paper is concerned with the further analysis and exploration of
the model introduced in
\cite{jort}, to which  we refer for a
more detailed and extensive discussion of the relevant literature
and of the motivation underlying the study of random heteropolymers
in general (e.g. their role as precursors of models for proteins),
and lattice gas models of random heteropolymers in
particular (see also e.g. \cite{mdoi,GOT,GOP} and references therein). 
We would like to emphasize that this model does not 
represent realistically the polymeric features of a 
heteropolymer, but it does capture some of the characteristics, 
and for this reason it can be considered as a precursor for 
a model of random heteropolymers. 
We complement the equilibrium study \cite{jort} in two distinct ways.
Firstly, we solve
the dynamics of the model (which we define as a stochastic process
of the Glauber type), by deriving exact closed
deterministic laws in the thermodynamics limit for a suitably
chosen set of dynamic order
parameters. This can be done, at least for site-disordered models,
as a direct consequence of the
simplification induced by the infinite dimensionality of the
model, and allows for an exact quantitative analysis of
the approach to equilibrium.
Secondly, we study the equilibrium properties of the model of
\cite{jort} in the limit of high local density.
Since the original
procedure followed in \cite{jort} for solving the model
(which applies to both site-disordered and bond-disordered
systems), does
not allow for this limit to be taken, we first present an
alternative (but equivalent) equilibrium solution (which applies
only to site disorder) where taking the high local density
limit is allowed. We then find, after an appropriate  re-scaling of the system
temperature (in order to prevent entropy dominance of the free
energy) a non-trivial phase diagram with a distinct swollen phase, a compact phase, and a phase characterised
by partial compactification of the heteropolymer.
Especially near the transition lines in this diagram, the system
is found to exhibit a highly non-trivial and slow plateau-type dynamics.

We supplement our theoretical analysis by extensive numerical simulations,
which support our predictions regarding the dynamics and regarding
the high-density phase diagram very satisfactorily (within the experimental
limitations imposed by the slow dynamics).

\section{Microscopic Rules of the Model}

The model of \cite{jort} describes a system of $N$ monomers, labeled
 $i\in\{1,..,N\}$, and $R=\alpha N$ possible sites for these monomers, labeled $r\in\{1,..,R\}$.
The degrees of freedom of the system are the locations $r_i\in\{1,\ldots,R\}$
of the $N$ monomers, and the overall system state is defined by the
state vector $\br=(r_1,\ldots,r_N)$.
Each site, however, can occupy at most $n_c$ monomers; this
imposes $R$ constraints on the state vector of the form $\sum_{i}\delta_{r_i,r}\leq n_c$
(one for each $r$), and limits the global
monomer  density according to $\alpha\geq n_c^{-1}$.
Those positions at a given site which are not occupied by monomers, are
assumed to be occupied by solvent molecules.
The monomers are allowed
to interact (in pairs) only when they find themselves at the same
site. Within equilibrium statistical mechanics this is described
by a Hamiltonian of the form
\be
H(\br)=\sum_{i<j} \lambda_{ij} \delta_{r_i,r_j}
\label{eq:hamiltonian}
\ee
The disorder in the system is in the realisation of the
interaction energies $\lambda_{ij}$.
Unlike \cite{jort}, we will in this paper restrict ourselves to
models with site disorder, where
\be
\lambda_{ij}=\phi(\lambda_i,\lambda_j)
\label{eq:site_disorder}
\ee
such as the Random Hydrophobic Model (RHM) \cite{Ob,GLO,TAB},
where $\phi(\lambda,\lambda^\prime)=\frac{1}{2}(\lambda+\lambda^\prime)$,
or the Random Charge Model (RCM) \cite{GO_C,Sf}, where
$\phi(\lambda,\lambda^\prime)=\epsilon \lambda_i \lambda_j$.
Here the variable $\lambda_i$ defines the monomer species of
monomer $i$, with $\lambda_{i}>0$
for hydrophylic/polar monomers, and $\lambda_{i}<0$ for hydrophobic monomers.
The RHM describes a system where hydrophobic monomers try to avoid contact with the
solvent, and all hydrophilic monomers try to
optimise contact with the solvent.
The parameter $\epsilon\in\{-1,1\}$ in the RCM controls whether monomers of
identical polarity attract or repel
each other.
We restrict ourselves, for simplicity, to the case whether the disorder is discrete in nature
and where there are $K$ monomer species,
i.e. $\lambda_{i} \in \{ \Lambda_1,...,\Lambda_{K}\}$.

We next endow this equilibrium system with a Glauber-type
dynamics. Time is discretised, and the probability to find system
state $\br$ at iteration step $\ell$ is written as $p_\ell(\br)$.
The elementary transitions we will allow for are those
where a single (randomly drawn) monomer is moved
to a (randomly drawn) new site;
to describe such transitions we define an operator $F_{ix}$ which moves
monomer $i$ to site $x$: $F_{ix}(\br)=(r_{1},...,r_{i-1},x,r_{i+1},...,r_{N})$.
The evolution of the system is a Markovian
stochastic process of the form
\be
p_{\ell+1}(\br)=\sum_{\br^\prime}W[\br;\br^\prime]p_{\ell}(\br^\prime)
\label{eq:Markov}
\ee
with the transition probabilities
\begin{eqnarray}
W[\br,\br^\prime]&=&\frac{1}{NR}\sum_{i=1}^N\sum_{x=1}^R
[\prod_{j\neq i}\delta_{r_j,r_j^\prime}]
\left\{
\delta_{n_x(\br^\prime),n_c}\delta_{r_i,r_i^\prime}~+
\right.
\label{eq:transitions}
 \\
&&\left.
\hspace*{-10mm}
\overline{\delta}_{n_x(\br^\prime),n_c}
\left[
p(F_{ix}\br^\prime;\br^\prime)\delta_{r_i,x}+[1\minus p(F_{ix}\br^\prime;\br^\prime)]\delta_{r_i,r_i^\prime}
\right]
\right\}
\nonumber
\end{eqnarray}
and
\begin{eqnarray}
p[\br;\br^\prime]&=&
\frac{e^{-\frac{1}{2}\beta\Delta H(\br;\br^{\prime})}}{2\cosh [\frac{1}{2}\beta
\Delta H(\br;\br^{\prime})]}
\label{eq:Glauber}
\end{eqnarray}
in which $\Delta H(\br;\br^{\prime})=H(\br)-H(\br^\prime)$,
 $\overline{\delta}_{ij}=1-\delta_{ij}$, and $n_x(\br)=\sum_i\delta_{r_i,x}$
(the number of monomers found at site $x$).
The process (\ref{eq:Markov}) is one where
we forbid all transitions which
would lead to violation of the constraint $n_x(\br)\leq n_c$, and where we
allow all other single particle moves
(where both the candidate particle $i$ to be moved and the candidate
site $x$ for it to move to are drawn uniformly at random)
with Glauber-type probabilities, such that the invariant distribution will be the Boltzmann state
(provided the system is ergodic, which in the present case is intuitively clear).
One easily convinces oneself that the transition matrix
(\ref{eq:transitions}) is properly normalised, i.e.
$\sum_{\br}W[\br,\br^\prime]=1$ for all $\br^\prime$, and that the
Markov chain (\ref{eq:Markov}) indeed obeys detailed balance, i.e. $W[\br,\br^\prime]p_\infty(\br^\prime)=
W[\br^\prime,\br]p_\infty(\br)$ for all $\{\br,\br^\prime\}$, with $p_\infty(\br)\sim \exp[-\beta
H(\br)]$.

Although the Markov chain (\ref{eq:Markov}) is the representation which is most conveniently
implemented in numerical simulations, for a theoretical analysis
of the dynamics a continuous time description would be preferable.
This is achieved (see \cite{Bedeaux}) upon choosing the duration
of each iteration step of (\ref{eq:Markov}) to be a Poisson
distributed random number (with average $\tau$), which converts the Markov chain into a
master equation of the form
\bd
\tau\frac{d}{dt}p_t(\br)
=\sum_{\br^{\prime}}W[\br;\br^{\prime}]p_t(\br^{\prime})- p_t(\br)
\ed
Upon choosing $\tau=N^{-1}$ (so that every monomer is targeted for
update on average once per unit time), and upon inserting (\ref{eq:transitions}), we then arrive at the master
equation
\begin{eqnarray}
\frac{d}{dt}
p_{t} (\br)&=& \frac{1}{R}\!
\sum_{i x}\sum_{\br^{\prime}}p_{t}(\br^{\prime})p[F_{ix}\br^{\prime};\br^{\prime}]
\nonumber \\
&&
\times~\overline{\delta}_{n_{x}(\br^{\prime}),n_{c}}
 \left[
\delta_{\br,F_{ix}\br^{\prime}}\minus
\delta_{\br,\br^{\prime}}
\right]
\label{eq:master}
\end{eqnarray}
Using the properties of the Poisson distribution, one can show
that the relative uncertainty on the time axis, induced by having random
durations of iteration steps, will vanish for $N\to\infty$.

We finally note that the only remaining
non-trivial Glauber term (\ref{eq:Glauber}) occurring
in (\ref{eq:master}) is of the following form:
\be
p[F_{ix}\br;\br]=\frac{1}{2}-\frac{1}{2}
\tanh[\frac{1}{2}\beta \Delta_{ix} H(\br)]
\label{eq:moves}
\ee
with $\Delta_{ix} H(\br)=H(F_{ix}\br)\minus
H(\br)$, which is here given by
\bd
\Delta_{ix} H(\br)=
\sum_{j} \overline{\delta}_{ij}\phi(\lambda_i,\lambda_j)\left[\delta_{x,r_j}\minus \delta_{r_i,r_j}\right]
\ed
Upon defining
$n_{\lambda,x}(\br)=\sum_i\delta_{r_i,x}\delta_{\lambda_i,\lambda}$
(counting the number of type-$\lambda$ monomers at site
$x$), we can write this as
\be
\Delta_{ix} H(\br)
=
\sum_{\lambda}\phi(\lambda_i,\lambda)
\left[n_{\lambda,x}(\br)
\minus n_{\lambda,r_i}(\br)\right]
+
\phi(\lambda_i,\lambda_i)\overline{\delta}_{x,r_i}
\label{eq:fullDeltaH}
\ee

\section{Dynamics: Derivation of Closed Macroscopic Laws}

The appropriate dynamic order parameters
of the system are the smallest set of macroscopic
observables which obey closed and deterministic
evolution equations in the thermodynamic limit.
In the present model this set turns out to be, at least for
finite $n_c$ and for pair energies of the separable form
(\ref{eq:site_disorder}), the fraction $c_{\bn}(\br)$ of sites with a given
local population (or `local state') $\bn=(n_{\Lambda_1},\ldots,n_{\Lambda_K})$ of monomers.
Here $n_\lambda$ indicates the number of monomers of species
$\lambda$.
Thus
\be
c_{\bn}(\br)=
\frac{1}{R} \sum_{x=1}^R \!\left[
\prod_{\lambda=\Lambda_{1}}^{\Lambda_K}
\delta_{n_{\lambda,x}(\br),n_{\lambda}} \right]
\label{eq:orderparameter}
\ee
with $\bn\in\{0,\ldots,n_c\}^K$. Note that $\sum_{\bn}c_{\bn}(\br)=1$
 and that
$c_{\bn}(\br)=0$ as soon as $\sum_{\lambda}n_\lambda>n_c$ (due to
our constraint on the maximum number of monomers at any given
site), for any $\br$. We will write the constraint as
$g[\bn]\in\{0,1\}$, where $g[\bn]=1$ if and only if  $\sum_{\lambda}n_\lambda \leq
n_c$.

The probability of finding the set of obervables $\{c_{\bn}(\br)\}$
at time $t$ is given by
\be
P_{t}[\{c_{\bn}\}]= \sum_{\br} p_{t}(\br) \prod_{\bn}\delta[c_{\bn}\minus c_{\bn}(\br)]
\label{eq:macrostpt}
\ee
Its evolution in time follows upon taking a
temporal derivative and inserting the master equation
(\ref{eq:master}):
\begin{eqnarray*}
\frac{d}{dt}
P_{t}[\{c_{\bn}\}]&=&\frac{1}{2R}
\sum_{\br}p_{t}(\br)\!\sum_{i x}
\left[1\minus \tanh[\frac{1}{2}\beta \Delta_{ix} H(\br)]\right]
\\
&&\hspace*{-20mm}
\times~\overline{\delta}_{n_{x}(\br),n_{c}}
\left\{
 \prod_{\bn}\delta[c_{\bn}\minus c_{\bn}(F_{ix}\br)]
-\prod_{\bn}\delta[c_{\bn}\minus c_{\bn}(\br)]
\right\}
\end{eqnarray*}
We expand this expression, using
\begin{eqnarray*}
c_{\bn}(F_{ix}\br)
&=&c_{\bn}(\br)
\\
&&
\hspace*{-15mm}
+\frac{1}{R}\overline{\delta}_{x,r_i}
[\delta_{n_{\lambda_i,r_i}(\br),n_{\lambda_i}+1}\minus \delta_{n_{\lambda_i,r_i}(\br),n_{\lambda_i}}]
\prod_{\lambda\neq \lambda_i}\delta_{n_{\lambda,r_i}(\br),n_{\lambda}}
\\
&&
\hspace*{-15mm}
+\frac{1}{R}\overline{\delta}_{x,r_i}
[\delta_{n_{\lambda_i,x}(\br),n_{\lambda_i}-1}\minus\delta_{n_{\lambda_i,x}(\br),n_{\lambda_i}}]
\prod_{\lambda\neq \lambda_i}\delta_{n_{\lambda,x}(\br),n_{\lambda}}
\end{eqnarray*}
We also insert the partitionings
$1=\sum_{\bn^\pprime}[\prod_{\lambda}
\delta_{n_{\lambda,x}(\br),n^\pprime_{\lambda}}]$ and
$1=\sum_{\bn^\ppprime}[\prod_{\lambda}
\delta_{n_{\lambda,r_i}(\br),n^\ppprime_{\lambda}}]$,
and we use
(\ref{eq:fullDeltaH}).
This then, together with the identity
$n_{x}(\br)=\sum_{\lambda} n_{\lambda,x}(\br)$,
leads to the following differential equation for $P_{t}[\{c_{\bn}\}]$:
\begin{eqnarray*}
\frac{d}{dt}
P_{t}[\{c_{\bn}\}]&=&
-\sum_{\bn^\prime}\frac{\partial}{\partial c_{\bn^\prime}}
\left\{
P_{t}[\{c_{\bn}\}]
\sum_{\lambda}\sum_{\bn^\pprime}\sum_{\bn^\ppprime}
c_{\bn^\pprime}
c_{\bn^\ppprime}
\right.
\\
&&
\hspace*{-17mm}
\left.
\times~\frac{1}{2}n^\ppprime_{\lambda}
\left[1\minus \tanh[\frac{1}{2}\beta[
\sum_{\lambda^\prime}\phi(\lambda,\lambda^\prime)
(n^\pprime_{\lambda^\prime}\minus n^\ppprime_{\lambda^\prime})
\plus \phi(\lambda,\lambda)]]
\right]
\right.
\\
&&
\hspace*{-17mm}
\left.
\times~
\overline{\delta}_{\sum_{\lambda^\prime}n^\pprime_{\lambda^\prime},n_{c}}
[
(\delta_{n^\ppprime_{\lambda},n_{\lambda}+1}\minus
\delta_{n^\ppprime_{\lambda},n_{\lambda}})
\!\prod_{\lambda^\prime\neq \lambda}\delta_{n^\ppprime_{\lambda^\prime},n_{\lambda^\prime}}
\right.
\\[-3mm]
&&\hspace*{-2mm}
\left.
+~
(\delta_{n^\pprime_{\lambda},n_{\lambda}-1}\minus\delta_{n^\pprime_{\lambda},n_{\lambda}})
\!\prod_{\lambda^\prime\neq \lambda}\delta_{n^\pprime_{\lambda^\prime},n_{\lambda^\prime}}
]\right\}
+\order(\frac{1}{N})
\end{eqnarray*}
Taking the limit $N\to\infty$ for finite $n_c$ and finite $K$ converts this into
a Liouville equation, describing deterministic flow, i.e.
solutions of the form
$P_{t}[\{c_{\bn}\}]=\prod_{\bn}\delta[c_{\bn}-c_{\bn}(t)]$
where the deterministic trajectories $\{c_{\bn}(t)\}$ are simply
given by the solution of the associated flow equations.
With the short-hands $F_\lambda^{\pm}\bn=(n_{\Lambda_1},\ldots,n_{\lambda-1},n_\lambda\pm 1,n_{\lambda+1},
\ldots,n_{\Lambda_K})$
these (coupled) flow
equations can be simplified to
\begin{eqnarray}
\frac{d}{dt}
c_{\bn}
&=&
\sum_\lambda \sum_{\bn^\prime}c_{\bn^\prime}\left\{
c_{F_\lambda^+\bn}(n_{\lambda}\plus 1)
\overline{\delta}_{\sum_{\mu}\!\!n^\prime_{\mu},n_{c}}
\varepsilon_\lambda[\bn^\prime\!\minus \bn]
\right.
\nonumber \\
&&
\left.
\hspace*{15mm}
+~
c_{F_\lambda^-\bn}
n^\prime_{\lambda}
\overline{\delta}_{\sum_{\mu}\!\!n_{\mu},n_{c}+1}
\varepsilon_\lambda[\bn\minus \bn^\prime]
\right.
\nonumber \\
&&
\left.
\hspace*{15mm}
-~
c_{\bn}n_{\lambda}
\overline{\delta}_{\sum_{\mu}\!\!n^\prime_{\mu},n_{c}}
\varepsilon_\lambda[F_\lambda^+\bn^\prime\minus\bn]
\right.
\nonumber \\
&&\left.
\hspace*{15mm}
-~
c_{\bn}n^\prime_{\lambda}
\overline{\delta}_{\sum_{\mu}\!\!n_{\mu},n_{c}}
\varepsilon_\lambda[\bn\minus F_\lambda^-\bn^\prime]
\right\}
\label{eq:macrodynamics}
\end{eqnarray}
where
\be
\varepsilon_\lambda[\bm]=
\frac{1}{2}
\minus \frac{1}{2}\tanh[\frac{1}{2}\beta
\sum_{\lambda^\prime}\phi(\lambda,\lambda^\prime)m_{\lambda^\prime}]
\label{eq:macrorates}
\ee
The imposed limit on the maximum number of monomers at a given site
is seen to be built into (\ref{eq:macrodynamics}), since for any local monomer
arrangement $\bn$ with $\sum_{\lambda}n_{\lambda}=n_c\plus 1$
equation (\ref{eq:macrodynamics}) indeed dictates that $\frac{d}{dt}c_{\bn}=0$
if $c_{\bn}=0$.

Since the dimensionality of our macroscopic laws increases with $n_c$
and the number $K$ of monomer types, further reductions can be helpful. One reduction is to consider
the fraction $p[n_+,n_-]$ of sites with local monomer arrangements
where $n_+$ monomers have $\lambda_i>0$ and $n_-$ monomers have $\lambda_i<0$
(for the RHM this implies counting hydrophobic and hydrophilic monomers; for the RCM one counts monomers
with a specific charge sign):
\be
p[n_+,n_-]=\sum_{\bn}c_{\bn}~\delta_{\sum_{\lambda>0}\!n_\lambda,n_+}~\delta_{\sum_{\lambda<0}\!n_\lambda,n_-}
\label{eq:plusminus}
\ee
with $n_\pm\in\{0,n_c\}$.
We will, more specifically, inspect the (co-)variances $\sigma^2_\pm =\bra
n_\pm^2\ket-\bra n_\pm\ket^2$
and $\rho=\bra n_+n_-\ket$.
They give information about the spatial properties of the
system state:
a large $\sigma_\pm$ implies spatial concentration of a monomer type in a relatively small number of sites,
 whereas $\rho$ informs us of the degree of spatial separation of
 the monomer types.
These quantities $p[n_+,n_-]$ will normally not obey closed
equations, but are calculated via (\ref{eq:plusminus}) from the (numerical) solution
of (\ref{eq:macrodynamics}).

\section{Dynamics: Comparison with Numerical Simulations}

\begin{figure}[t]
\vspace*{10mm}
\setlength{\unitlength}{0.55mm}
\begin{picture}(0,100)
\put( -15,10){\epsfxsize=140\unitlength\epsfbox{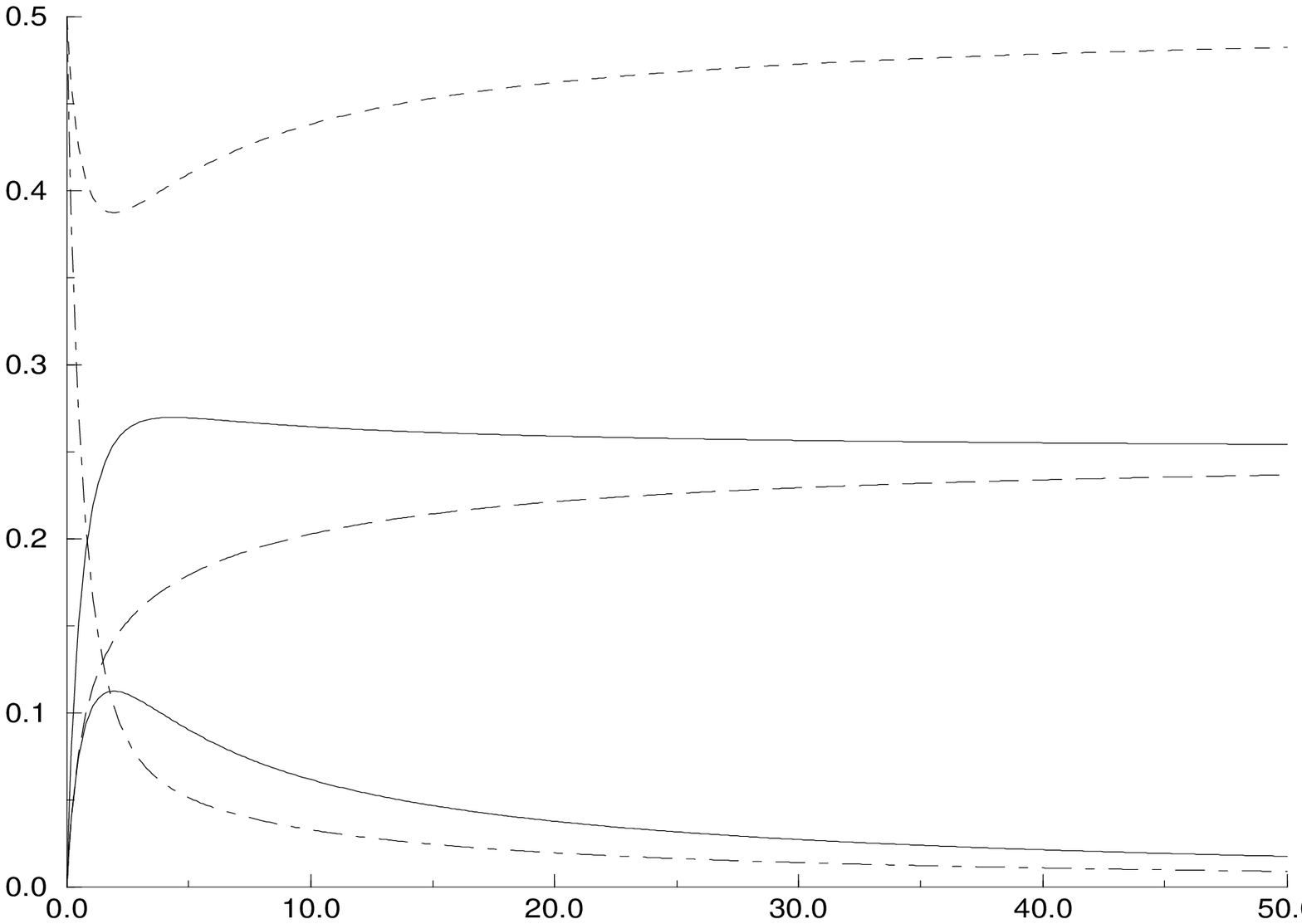}}
\put(65,105){\makebox(0,-5){$c_{(1,0)}$ }}
\put(65,80){\makebox(0,-5){$c_{(0,0)}$ }}
\put(65,65){\makebox(0,-5){$c_{(0,2)}$ }}
\put(65,43){\makebox(0,-5){$c_{(1,1)}$ }}
\put(20,35){\makebox(0,-5){$c_{(0,1)}$ }}
\put(73,21){\makebox(0,-5){\large $t$}}
\put(-15,-102){\epsfxsize=140\unitlength\epsfbox{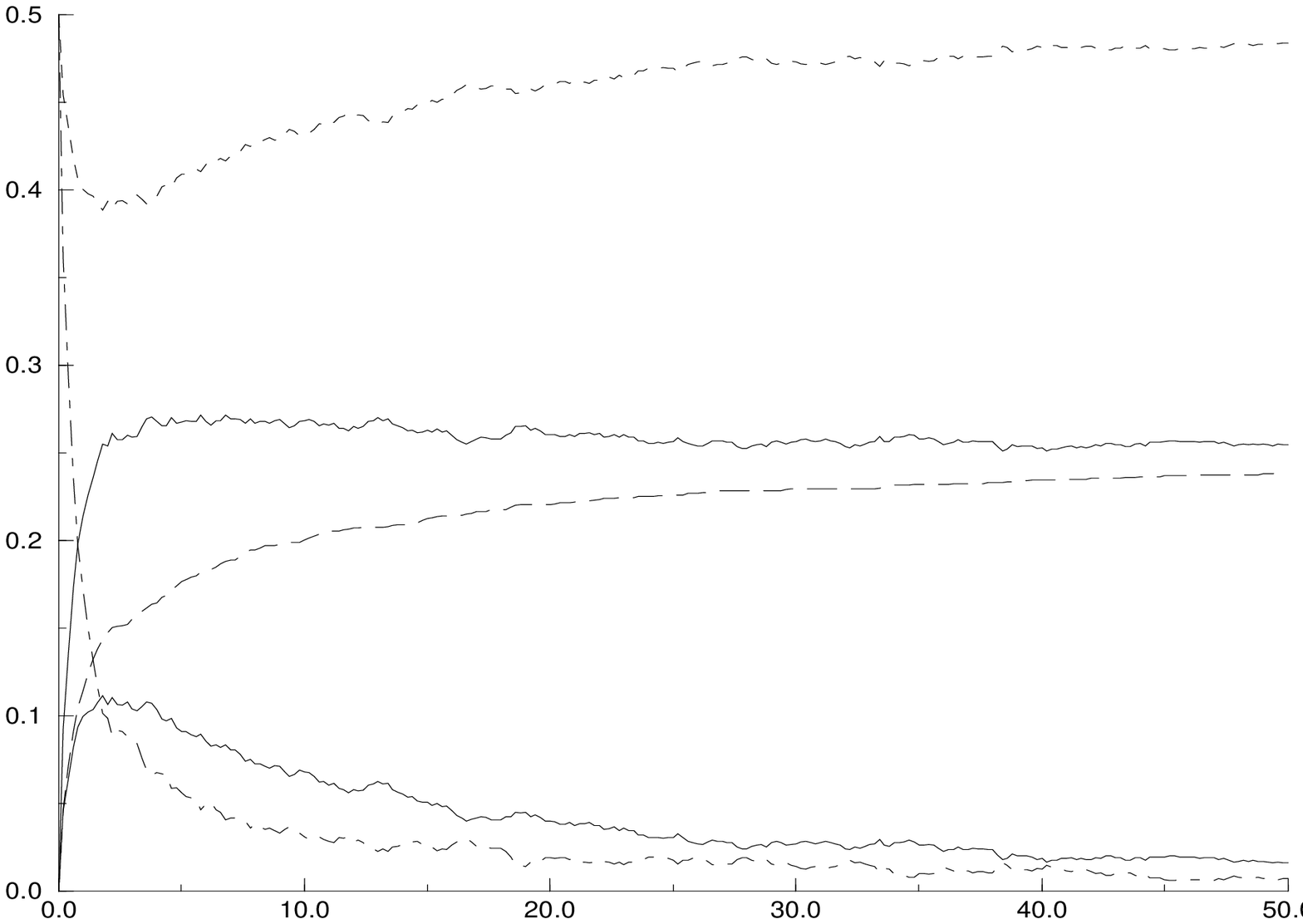}}
\put(65,-7){\makebox(0,-5){$c_{(1,0)}$ }}
\put(65,-32){\makebox(0,-5){$c_{(0,0)}$ }}
\put(65,-47){\makebox(0,-5){$c_{(0,2)}$ }}
\put(65,-69){\makebox(0,-5){$c_{(1,1)}$ }}
\put(20,-77){\makebox(0,-5){$c_{(0,1)}$ }}
\put(73,-91){\makebox(0,-5){\large $t$}}
\end{picture}
\vspace*{55mm}
\caption{
Evolution of the order parameters $\{c_{\bn}\}$ for the Random
Hydrophilic Model (RHM), with $K=2$ (two monomer types,
$\lambda\in\{-1,1\}$,
one hydrophobic and one
hydrophilic), $n_{c}=2$, $\alpha=1$, and $\beta=100000$.
Upper graphs: theory, i.e. numerical solution of (\ref{eq:macrodynamics}).
Lower graphs: numerical simulations, with $N=3000$.
Note: in this example $c_{(2,0)}=0$ for all $t$.}
\label{fig:dynamicgraph}
\end{figure}

\begin{figure}[t]
\vspace*{10mm}
\setlength{\unitlength}{0.55mm}
\begin{picture}(0,100)
\put( -15,10){\epsfxsize=140\unitlength\epsfbox{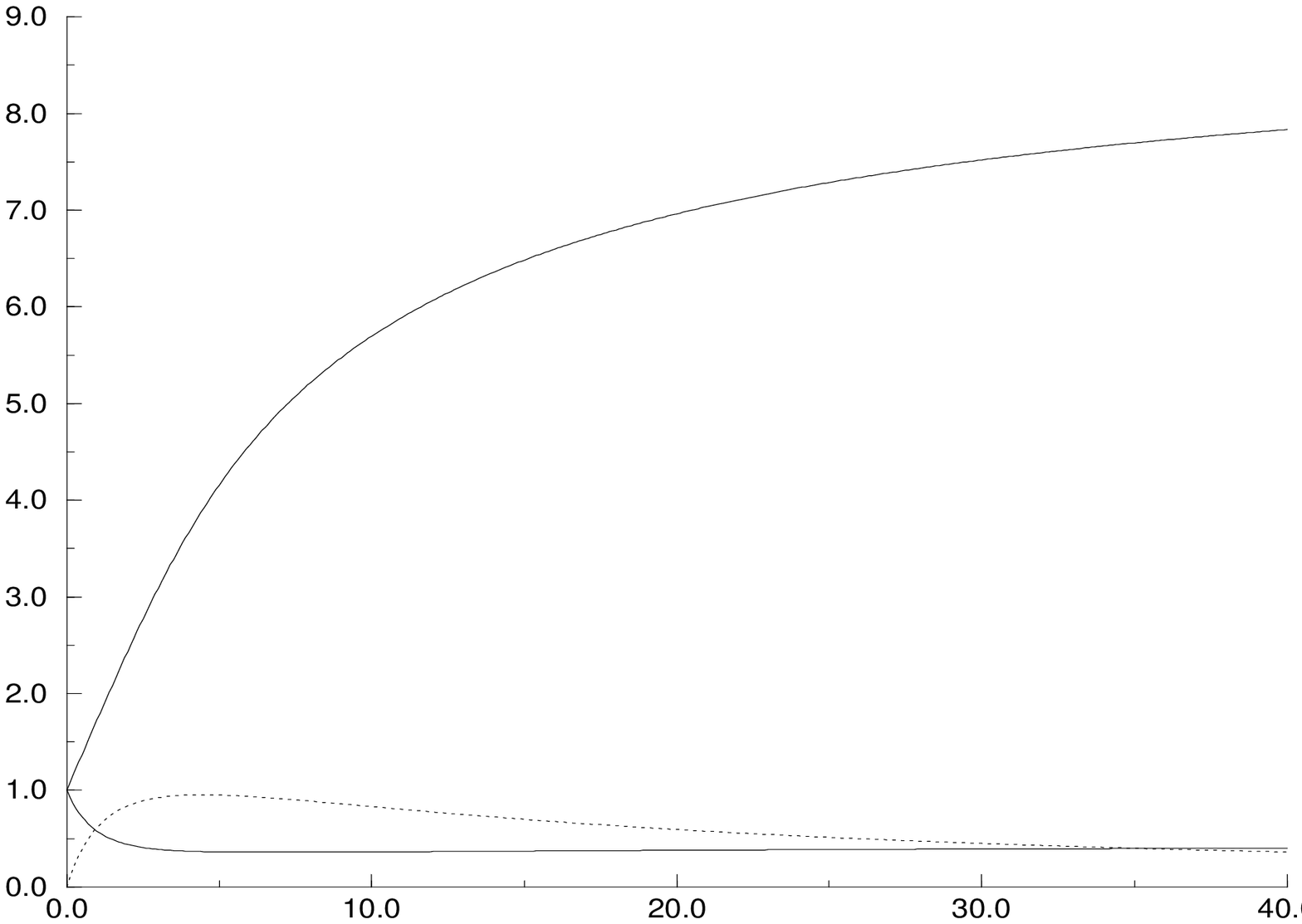}}
\put(73,21){\makebox(0,-5){\large $t$}}
\put(85,0){\makebox(0,-5){$\sigma^2_-$}}
\put(40,-67){\makebox(0,-5){$\rho$}}
\put(120,-72){\makebox(0,-5){$\sigma^2_+$}}
\put( -15,-102){\epsfxsize=140\unitlength\epsfbox{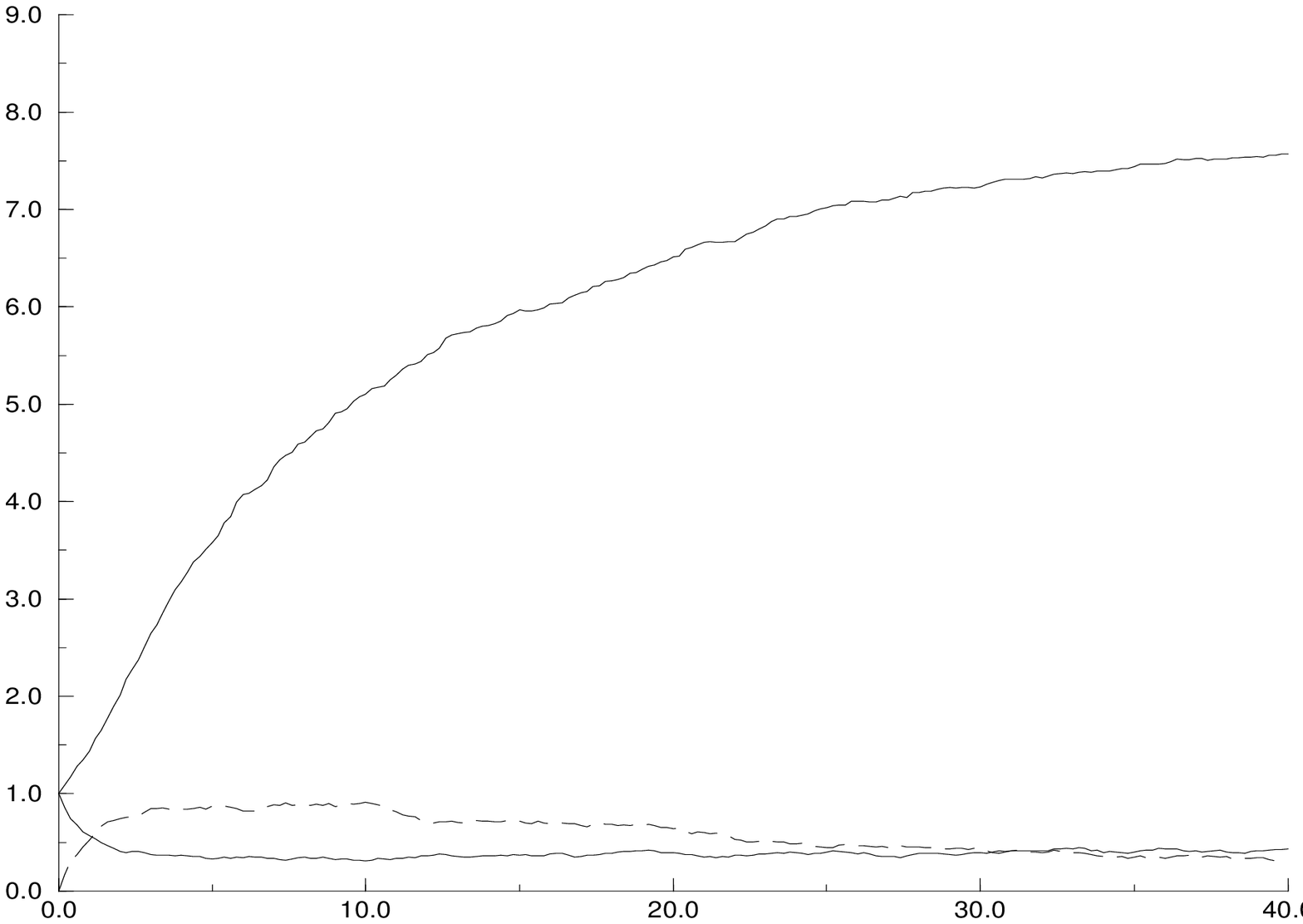}}
\put(73,-91){\makebox(0,-5){\large $t$}}
\put(85,112){\makebox(0,-5){$\sigma^2_-$}}
\put(40,45){\makebox(0,-5){$\rho$}}
\put(120,40){\makebox(0,-5){$\sigma^2_+$}}
\end{picture}
\vspace*{55mm}
\caption{
Evolution of the (co-)variances $\sigma_\pm^2=
\bra n_\pm^2\ket$ and $\rho=\bra n_+ n_-\ket-\bra n_+\ket\bra
n_-\ket$,
for the Random Hydrophilic Model (RHM), with $K=2$ ($\lambda\in\{-1,1\}$), $n_c=10$, $\alpha=\frac{1}{2}$, and $\beta=1$.
Initial conditions: $c_{(2,0)}=c_{(0,2)}=0.5$ (perfectly separated monomer species).
Upper graphs: theory, i.e. numerical solution of (\ref{eq:macrodynamics}).
Lower graphs: numerical simulations, with $N=2000$.
From top to bottom in both pictures: $\sigma^2_-(t)$ (with $\sigma^2_-(0)=1$),
$\rho(t)$ (dotted/dashed, with $\rho(0)=0$) and $\sigma^2_+(t)$
(with $\sigma^2_+(0)=1$).
}
\label{fig:dynavar1}
\end{figure}

\begin{figure}[t]
\vspace*{10mm}
\setlength{\unitlength}{0.55mm}
\begin{picture}(0,100)
\put( -15,10){\epsfxsize=140\unitlength\epsfbox{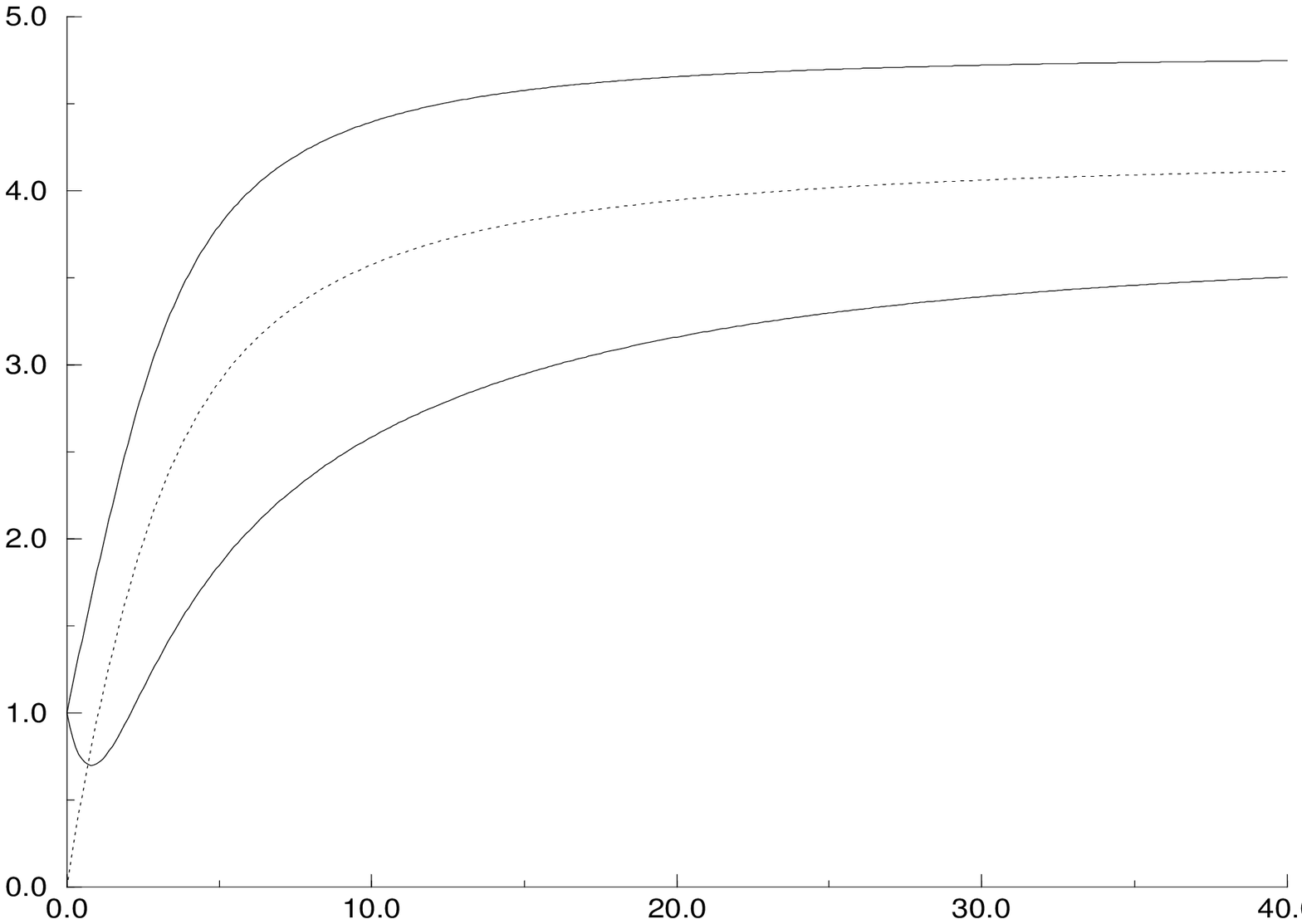}}
\put(73,21){\makebox(0,-5){\large $t$}}
\put(25,1){\makebox(0,-5){$\sigma^2_-$}}
\put(40,-17){\makebox(0,-5){$\rho$}}
\put(60,-36){\makebox(0,-5){$\sigma^2_+$}}
\put( -15,-102){\epsfxsize=140\unitlength\epsfbox{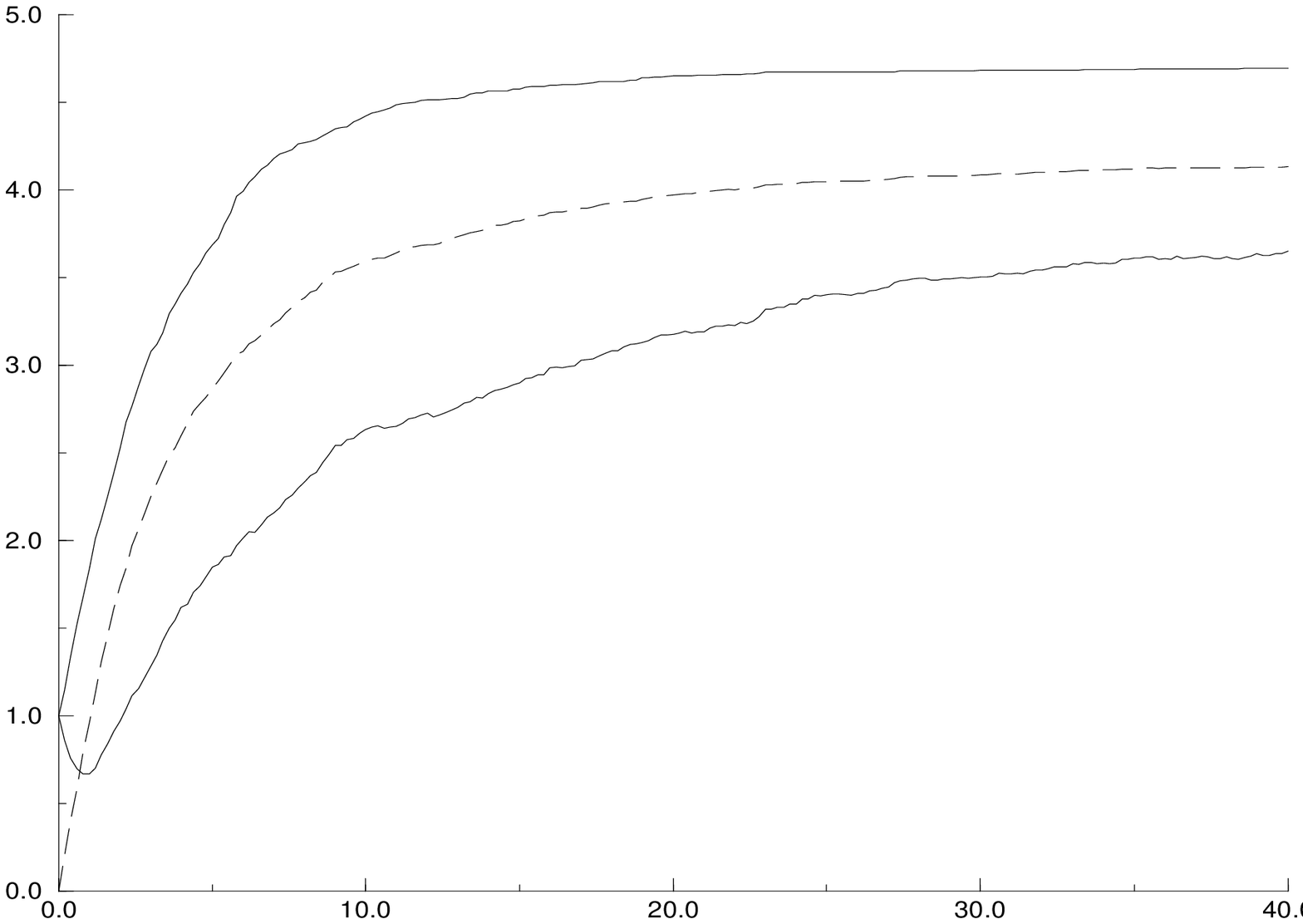}}
\put(73,-91){\makebox(0,-5){\large $t$}}
\put(25,113){\makebox(0,-5){$\sigma^2_-$}}
\put(40,95){\makebox(0,-5){$\rho$}}
\put(60,76){\makebox(0,-5){$\sigma^2_+$}}
\end{picture}
\vspace*{55mm}
\caption{
Evolution of the (co-)variances $\sigma_\pm^2=
\bra n_\pm^2\ket$ and $\rho=\bra n_+ n_-\ket-\bra n_+\ket\bra
n_-\ket$, for the Random Hydrophilic Model (RHM), with $K=2$ ($\lambda\in\{-10,1\}$, i.e. very strong hydrophobicity), $n_c=10$, $\alpha=\frac{1}{2}$, and $\beta=1$.
Initial conditions: $c_{(2,0)}=c_{(0,2)}=0.5$ (perfectly separated monomer species).
Upper graphs: theory, i.e. numerical solution of (\ref{eq:macrodynamics}).
Lower graphs: numerical simulations, with $N=2000$.
From top to bottom in both pictures: $\sigma^2_-(t)$ (with $\sigma^2_-(0)=1$),
$\rho(t)$ (dotted/dashed with $\rho(0)=0$) and $\sigma^2_+(t)$
(with $\sigma^2_+(0)=1$).
}
\label{fig:dynavar2}
\end{figure}

We first compare the numerical solution of the macroscopic laws (\ref{eq:macrodynamics})
with the results of measuring the observables
(\ref{eq:orderparameter}) directly in numerical simulations of the
microscopic process (\ref{eq:Markov},\ref{eq:transitions}). All simulations 
are presented as single runs, as oppose to ensemble averages.
A typical example is shown in figure \ref{fig:dynamicgraph}, for the RHM
(i.e. $\phi(\lambda,\lambda^\prime)=\frac{1}{2}(\lambda+\lambda^\prime)$, with $K=2$ (two monomer types,
$\lambda\in\{-1,1\}$, one hydrophobic and one
hydrophilic), $n_{c}=2$, $\alpha=1$ and $\beta=100000$.
In this case $\bn\in\{(0,0),(1,0),(0,1),(1,1),(2,0),\\ (0,2)\}$.
Note that $\sum_{\bn}c_{\bn}=1$ at all times, by definition.
We observe excellent agreement between theory and experiment.
The behaviour of the system can be understood as follows.
Hydrophilic monomers, with $\lambda=1$, prefer local arrangements
where contact with the solvent is maximised, i.e. where $n_+\in\{0,1\}$
(since the maximum number of molecules per site is $n_c=2$).
Hydrophobic monomers prefer the opposite, i.e. arrangement with $n_-\in\{0,2\}$.
Hence the observed evolution of the system towards a state
where only $c_{(0,0)}$, $c_{(0,2)}$ and $c_{(1,0)}$ are nonzero.

Note that as soon as $K=2$, with one hydrophobic and one
hydrophilic monomer species (as in the example above), the
observables in (\ref{eq:plusminus}) simply reduce to
$p[n_+,n_-]=c_{(n_+,n_-)}$, and the (co-)variances of $p[n_+,n_-]$
reduce to those of the $c_{\bn}$.
In figure (\ref{fig:dynavar1}) we show a first example of the
evolution of $\sigma^2_\pm=\bra n_\pm^2\ket$ and $\rho=\bra n_+ n_-\ket$,
for the RHM with $\beta=1$, $n_c=10$ and $\alpha=\frac{1}{2}$.
Here $\lambda_+=1$ and $\lambda_-=-1$, and the initial conditions
were $c_{(2,0)}=c_{(0,2)}=0.5$ (i.e. perfectly separated monomer species).
We observe again a perfect agreement between theory and
(numerical) experiment.
The development of a large variance $\sigma_-$ for the hydrophobic monomers
is explained by the observation that such
monomers promote arrangements of the type $(n_+,n_-)=(0,n_c)$
or $(n_+,n_-)=(\star,0)$ (i.e. $n_-\in\{0,n_c\}$, without
intermediate values).
Hydrophilic monomers, on the other hand, promote local arrangements of the type
$(n_+,n_-)=(1,\star)$,
i.e. with a single optimal value for $n_+$, hence the low variance
$\sigma_+$. The low observed correlation $\rho$ indicates that the
various objectives can be achieved independently.
The situation is quite different for the system illustrated in figure
(\ref{fig:dynavar2}), which only differs from the previous one in
that the hydrophobicity of the hydrophobic monomers
has been strengthened from $\lambda=-1$ to
$\lambda=-10$.
Now the system will find it energetically
favourable to `absorb' hydrophilic monomers into sites where
hydrophobic  monomers are in contact with solvent (that is, in
addition to
the system aiming for its hydrophobic monomers to be in $(n_+,n_-)=(0,n_c)$
arrangements, wherever possible).
The energetic gain of the hydrophobic monomers, which results upon having their hydrophilic cousins
replacing solvent molecules,
overrules the
 energetic loss of the hydrophilic ones, when the latter are taken away from solvent and put into contact with
hydrophobic monomers.
The primary aim of the hydrophobic
monomers is to avoid water molecules, and it is acceptable for this to be achieved by
sharing sites with hydrophilic monomers. Hence, virtually
any proposed move is accepted which involves a monomer going to a site with an adequate number of
hydrophobic monomers, regardless of the species of the
mobile monomer.
As a result we observe, in addition to the increase of the variance $\sigma^2_-$ of the hydrophobic monomers,
an increase of both the variance $\sigma^2_+$ of the hydrophilic monomers
and of the covariance $\rho$.

\section{Statics: An Alternative Solution}

In the remainder of this paper we will concentrate on
transitions between low density (swollen) and high density (collapsed) states.
We distinguish between the overall density $\alpha^{-1}=N/p$ (which
will remain finite) and the local density as limited by $n_c$, which is the quantity
which will be allowed to diverge.
In \cite{jort} the (exact) equilibrium solution of site-disordered models
(\ref{eq:hamiltonian},\ref{eq:site_disorder}) was derived
in a form which was specifically convenient to deal with many-valued or continuous disorder distributions;
for large $n_c$, however, it is very cumbersome (taking the $n_c\to\infty$ limit
would be ruled out).
Since finding collapsed states requires $n_c\to \infty$, we will first give an alternative solution method, which is
complementary to that of \cite{jort} in that it is convenient for large $n_c$,
but inappropriate to deal with
many-valued disorder
distributions.
In addition it will enable a natural and transparent connection with the dynamical
solution (\ref{eq:macrodynamics}).
We write the partition function for the system (\ref{eq:hamiltonian},\ref{eq:site_disorder}) as
\be
\cZ=\sum_{\br}\left[ \prod_{x=1}^R \theta_{n_c,n_x(\br)}\right]
e^{-\beta \sum_{i<j}\phi(\lambda_i,\lambda_j)\delta_{r_i,r_j}}
\label{free_energy}
\ee
with the short-hand
\[
\theta_{x,y}=\left\{ \begin{array}{lllll} 1&&&&{\rm for}~~ x>y\\ 0&&&&{\rm for}~~ x\leq y\end{array}
\right.
\]
(and its obvious vectorial generalisation
$\theta_{\bx,\by}=\prod_i\theta_{x_i,y_i}$).
We introduce the vectors $\bm=(m_1,\ldots,m_R)$ with $m_x\in\{0,\ldots,n_c\}$
to indicate the total number of monomers found at each of the $R$
sites. Similarly we define the vector $\bm_c=(n_c,\ldots,n_c)$.
This allows us to write the system's free energy as
$F=-\beta^{-1}\log \sum_{\bm}\theta_{\bm_c,\bm} \cZ[\bm]$, where the
`conditioned' partition functions $\cZ[\bm]$ are given by
\be
\cZ[\bm ]=\sum_{\br }\left[ \prod_{x=1}^R \delta_{m_x,n_x(\br)}\right]
e^{-\beta \sum_{i<j}\phi(\lambda_i,\lambda_j)\delta_{r_i,r_j}}
\label{eq:F_highd}
\ee
We consider only the scaling regime where the limit $n_c\to \infty$ is taken {\em after}
the thermodynamic limit $N\to\infty$. This allows us to solve our model in terms of the familiar sub-lattices
$I_\lambda=\{i|~\lambda_i=\lambda\}$, of size $|I_\lambda|$,
and their associated local occupation numbers $n_{\lambda,x}(\br)=\sum_{i\in
I_\lambda}\delta_{x,r_i}$. The latter are, in turn, also grouped into $R$-dimensional vectors
according to
$\bm_{\lambda}(\br)=(n_{\lambda,x}(\br),\ldots,n_{\lambda,R}(\br))$,
which allows us to write the conditioned partition function as
\begin{eqnarray}
\cZ[\bm]&=&
e^{{\frac{1}{2}\beta}\sum_i\phi(\lambda_i,\lambda_i)}\prod_\lambda \left[
\sum_{\bm_\lambda} {\cal N}_{|I_\lambda|}[\bm_\lambda ]\right]
\nonumber \\
& &
\times~\delta_{\bm,\sum_\lambda\! \bm_\lambda }
e^{-\frac{1}{2}\beta\sum_{\lambda,\lambda^\prime}\phi(\lambda,\lambda^\prime)\bm_\lambda
\cdot\bm_{\lambda^\prime}}
\end{eqnarray}
where
\bd
{\cal N}_M[\bm ]={M!\over m_1!..m_R!}~~\delta_{\sum_x m_x,M}
\ed
is a combinatorial term, counting the number of ways in which $M$ objects can be grouped into
$R$ subsets of prescribed sizes $\{m_1,..,m_R\}$.
We note that $\lim_{N\to\infty}|I_\lambda|/N=p_\lambda$
(the a priori probability to draw a monomer of type $\lambda$ randomly).
In the thermodynamic limit $N\to\infty$
the free energy per monomer $f=\lim_{N\to\infty}F/N$ becomes
\begin{eqnarray}
f&=&
-\frac{1}{2}\sum_\lambda p_\lambda\phi(\lambda,\lambda)
\nonumber \\
&&-\lim_{N\to\infty}\frac{1}{\beta N}\log
\sum_{\bm}\theta_{\bm_c,\bm}
\sum_{\{\bm_\lambda\}}
\delta_{\bm,\sum_\lambda\! \bm_\lambda }
\nonumber \\
&&
\prod_\lambda\! \left[
{\cal N}_{|I_\lambda|}[\bm_\lambda ]\right]
e^{-\frac{1}{2}\beta\sum_{\lambda,\lambda^\prime}\phi(\lambda,\lambda^\prime)\bm_\lambda
\cdot\bm_{\lambda^\prime}}
\end{eqnarray}
Introducing integral representations for the $\delta$-constraints  in
the terms ${\cal N}_{|I_\lambda|}[\bm_\lambda ]$, i.e.
$\delta_{nm}=(2\pi)^{-1}\int_0^{2\pi}\!du~e^{iu(n-m)}$, leads to a full factorisation
over sites, and consequently to an expression to be evaluated by saddle-point
integration (provided $K$ remains finite as $N\to\infty$). This then allows us
to write the non-constant part of the free energy per monomer as:
\begin{eqnarray}
f&=&
-\frac{1}{\beta }\extr_{~\by}
\left\{
\alpha \log \left[\sum_{\{n_\lambda\}}
\frac{\theta_{n_c,\sum_\lambda\! n_\lambda}}{\prod_\lambda n_\lambda !}
\right.\right.
\nonumber \\
&&\left.\left.\hspace*{-5mm}
e^{\sum_\lambda y_\lambda n_\lambda
-\frac{1}{2}\beta\sum_{\lambda,\lambda^\prime}\phi(\lambda,\lambda^\prime)n_\lambda
n_{\lambda^\prime}}
\right]
-\sum_\lambda p_\lambda y_\lambda
\right\}
\label{eq:equil_solution}
\end{eqnarray}
Variation with respect to $\by=\{y_\lambda\}$ gives the
$K$ saddle-point equations
\be
\frac{p_\mu}{\alpha}=
\frac{1}{Z}
\sum_{\{n_\lambda\}}
\frac{n_\mu\theta_{n_c,\sum_\lambda\! n_\lambda}}{\prod_\lambda n_\lambda !}
e^{\sum_\lambda y_\lambda n_\lambda
-\frac{1}{2}\beta\sum_{\lambda,\lambda^\prime}\phi(\lambda,\lambda^\prime)n_\lambda
n_{\lambda^\prime}}
\label{eq:saddle}
\ee
with the normalisation factor
\be
Z=\sum_{\{n_\lambda\}}
\frac{\theta_{n_c,\sum_\lambda\! n_\lambda}}{\prod_\lambda n_\lambda !}
~e^{\sum_\lambda y_\lambda n_\lambda
-\frac{1}{2}\beta\sum_{\lambda,\lambda^\prime}\phi(\lambda,\lambda^\prime)n_\lambda
n_{\lambda^\prime}}
\label{eq:local_Z}
\ee
The relation of this equilibrium solution (\ref{eq:equil_solution},\ref{eq:saddle})
to physical observables and to the previously obtained dynamical equations
(\ref{eq:macrodynamics}) becomes clear if one adds a
suitable infinitesimal generating term to the Hamiltonian
(\ref{eq:hamiltonian}):
\bd
H(\br)\to H(\br)+\frac{\chi}{R}\sum_{x=1}^R \!\left[
\prod_{\lambda=\Lambda_{1}}^{\Lambda_K}
\delta_{n_{\lambda,x}(\br),n_{\lambda}} \right]
\ed
We repeat the above derivation for the extended Hamiltonian
and use the identity $\lim_{\chi\to 0}\partial f/\partial\chi=\bra
c_{\bn}(\br)\ket$,
where $\bra\ldots\ket$ denotes the thermal average in the
Boltzmann state, and where the observables $c_{\bn}(\br)$ are
defined in (\ref{eq:orderparameter}). This reveals that the
numbers $\bn=\{n_\lambda\}$ in (\ref{eq:equil_solution},\ref{eq:saddle})
are identical to the local occupation numbers in
(\ref{eq:macrodynamics}), and that
in equilibrium
\be
c_{\bn}=
\frac{1}{Z}
\frac{\theta_{n_c,\sum_\lambda\! n_\lambda}}{\prod_\lambda n_\lambda !}
~e^{\sum_\lambda y_\lambda n_\lambda
-\frac{1}{2}\beta\sum_{\lambda,\lambda^\prime}\phi(\lambda,\lambda^\prime)n_\lambda
n_{\lambda^\prime}}
\label{eq:stationarysoln}
\ee
Hence the saddle-point equations (\ref{eq:saddle}) simply state that
$\bra n_\mu\ket=p_\mu/ \alpha$. Furthermore, one easily confirms  that equation
(\ref{eq:stationarysoln}), together with (\ref{eq:saddle}), indeed
gives the stationary solution of the macroscopic dynamical laws (\ref{eq:macrodynamics}).

\section{Two Simple Models in the Limit $n_c\to\infty$}

We now turn to the simplest case, where there is only one type of monomer in
the system, such that a local arrangement at a given site is fully specified by the number of monomers
present, i.e. $\bn\to n$. Note also that just one
interaction energy $\phi=\phi(\lambda,\lambda)$ will be left in our equations.
Model types will thus only differ in terms of the interpretation given to the possible
values of $\phi$, e.g. for the RHM one has $\phi=\lambda$ ($\phi>0$ for hydrophilic monomers, $\phi<0$
for hydrophobic monomers), whereas for the RHM one has $\phi=\epsilon \lambda^2$
($\phi>0$ for mutually repelling monomers, $\phi<0$ for mutually attracting
monomers).
In this section we will for simplicity use the RHM terminology,
and refer to $\phi>0$ and $\phi<0$ as hydrophilic and hydrophobic
systems, respectively.
Equation (\ref{eq:stationarysoln}) now becomes
\be
c_n={                {1\over n!}~e^{ny-{\frac{1}{2}\beta}\phi n^2}\over
     \sum_{m=0}^{n_c}{1\over m!}~e^{my-{\frac{1}{2}\beta}\phi m^2}}
\label{eq:reducedstatsoln}
\ee
where $c_n$ gives  the fraction of sites with $n$ monomers, and where
$y$ is to be solved from the non-linear equation
(\ref{eq:saddle}), which reduces to
\be
{1\over \alpha}=\sum_{n=0}^{n_c}n~c_n
\label{eq:alpha}
\ee
This equation can also be written as
\be
\frac{1}{\alpha}=\frac{\partial}{\partial y}\log Z(y)
~~~~~~~~
Z(y)=\sum_{n=0}^{n_c}{1\over n!}~e^{ny-\frac{1}{2}\beta\phi n^2}
\label{eq:alternative}
\ee
The associated value of the free energy per monomer
(\ref{eq:equil_solution}) is given by
\be
-\beta f=
\alpha \log Z(y)
-y
\label{eq:simple_f}
\ee
We will next study the solution $y$ of
(\ref{eq:reducedstatsoln},\ref{eq:alpha})
and the monomer density
variance $\sigma^2=\bra n^2\ket-\bra n\ket^2=\partial^2\log Z/\partial y^2$
in the different temperature regimes, for large $n_c$.
Note that from $\partial^2\log Z/\partial y^2=\sigma^2\geq 0$ it also follows
that the solution of (\ref{eq:alternative}), and thus also of
(\ref{eq:reducedstatsoln},\ref{eq:alpha}),
is unique.
\vsp

In the high-temperature region, both cases (hydrophilic and hydrophobic) can be treated on the same
footing.
Expansion of (\ref{eq:alternative}) in powers of $\beta$ yields
\be
Z(y)=\left[1-\frac{1}{2}\beta\phi\frac{\partial^2}{\partial y^2}+\order(\beta^2)\right]
\sum_{n=0}^{n_c}\frac{e^{ny}}{n!}
\ee
We take the limit $n_c\to\infty$ and insert the result into
(\ref{eq:alternative}), which leads to the following quadratic equation for
$x\equiv e^y$:
\be
\alpha^{-1}=x-\frac{1}{2}\beta\phi (x+2x^2)+\order(\beta^2)
\ee
The solution gives
\be
y=-\log \alpha+\frac{1}{2}\beta \phi(1\plus\frac{2}{\alpha}) +\order(\beta^2)
\label{eq:sol_0}
\ee
such that
\be
\si^2 = {1\over\alpha}\left[ 1-\frac{\phi}{\alpha T}\right]
+\order(T^{-2})~~~~~~(T\to\infty)
\label{eq:var_0}
\ee
Hence, at finite overall density (i.e. for finite $\al$) and for $T\to \infty$, the
monomer density variance $\sigma$ always remains finite (so the system is in a swollen
state), due to entropy dominance.
\vsp

In the low temperature regime, on the other hand,
it will be clear that the hydrophilic ($\phi>0$) and the hydrophobic ($\phi<0$) models
could exhibit qualitatively different behaviour.
More specifically, for models with just one monomer type as studied in this section,
we can write the Hamiltonian (\ref{eq:hamiltonian}) as
\begin{eqnarray}
H(\br)&=&\frac{1}{2}\phi\sum_{i\neq j}\delta_{r_i,r_j}
=\frac{1}{2}\alpha N\phi[\frac{1}{R}\sum_x n^2_x(\br)-\frac{1}{\alpha}]
\nonumber \\
&=&\frac{1}{2}\alpha N\phi[\sigma^2+
\frac{1}{\alpha}(\frac{1}{\alpha}\minus 1)]
\label{eq:energy_variance}
\end{eqnarray}
Thus the ground state of the hydrophilic system is the one with minimal monomer
density variance, and that of the hydrophobic system is the
one with maximal variance. This allows us to determine the ground states
without effort, which provides a welcome test of our general equations
(\ref{eq:reducedstatsoln},\ref{eq:alpha}).
Extremisation of $\sigma^2$ by variation of the $\{c_n\}$, with two
Lagrange parameters for the constraints
$\sum_nc_n=1$ and $\sum_n n c_n=\alpha^{-1}$, shows that an extremum will have at most two $n$
with $c_n>0$. However, since according to the theory we are not allowed to vary the $\{c_n\}$
at will (they must be of the form (\ref{eq:reducedstatsoln})), any
extremum thus obtained must be shown to correspond to a specific
value of the order parameter $y$ in (\ref{eq:reducedstatsoln}).

We first analyse the ground state of the hydrophilic model, for $n_c\to\infty$. In order to
minimize the energy the system must minimise the sum
$\sum_{i<j}\delta_{r_i,r_j}$, so the monomers must spread out maximally over the
available sites (i.e. the distribution $\{c_n\}$ must be as narrow as possible), given the constraints
$\sum_nc_n=1$ and $\sum_n n c_n=\alpha^{-1}$.
We define $n_\alpha\in \{0,1,2,\ldots\}$ such that $n_\alpha\leq
\alpha^{-1}<n_\alpha+1$. The minimum energy arrangement of the $\{c_n\}$ can now
be written as
\begin{eqnarray}
\ell=n_\alpha:~~~~   &&~
c_{\ell}=1\minus \alpha^{-1}\plus n_\alpha
\label{eq:phil_1}
\\
\ell=n_\alpha+1: &&~
c_{\ell}=\alpha^{-1}\minus n_\alpha
\label{eq:phil_2}
\end{eqnarray}
with $c_\ell=0$ for all other $\ell$.
For the variance we find
\be
\sigma^2=
(\alpha^{-1}\minus n_\alpha)(1\plus n_\alpha\minus \alpha^{-1})
\label{eq:var_i_i}
\ee
The variance remains finite, and is zero when $\alpha^{-1}$
is an integer (in which case a state exists where all sites have equal
monomer numbers).
The system will be in a swollen state at any temperature;
there are no phase transitions, not even for $n_c\to\infty$.
We now show how this state can be extracted from (\ref{eq:reducedstatsoln},\ref{eq:alpha})
for $\beta\to\infty$.
It will be clear from (\ref{eq:alpha})
that a finite $\alpha$ solution requires $y^{-1}=\order(T)$ as $T\to
0$. Putting $y=\beta\phi x$ converts (\ref{eq:alpha})
into
\be
\frac{1}{\alpha}=\frac{\sum_{n\geq 0}{n\over n!}~ e^{-\frac{1}{2}\beta\phi(n-x)^2}}
{\sum_{n\geq 0}{1\over n!}~e^{-\frac{1}{2}\beta\phi(n-x)^2}}
\label{eq:intermediate1}
\ee
For $\beta\to\infty$ the two terms $n=n_x$ and $n=n_x\plus 1$ will dominate
the summations in (\ref{eq:intermediate1}), where
$n_x\in \{0,1,2,\ldots\}$ such that $n_x\leq x<n_x+1$,
and
we obtain
\bd
\frac{1}{\alpha}=\lim_{\beta\to\infty}
\frac
{
n_x+
e^{-\beta\phi[n_x-x+\frac{1}{2}]}}
{
1+\frac{1}{n_x+1}~
e^{-\beta\phi[n_x-x+\frac{1}{2}]}
}
\ed
Upon making the ansatz $x=n_\alpha+\frac{1}{2}+\log(z)/\beta\phi$,
where
$n_\alpha\in \{0,1,2,\ldots\}$ such that $n_\alpha\leq
\alpha^{-1}<n_\alpha+1$,
and subsequently taking the $\beta\to\infty$ limit, we find:
\bd
\frac{1}{\alpha}=
\frac{[n_\alpha+\lim_{\beta\to\infty}z][n_\alpha+1]}
{n_\alpha+1+\lim_{\beta\to\infty}z}
\ed
with solution
\be
\lim_{\beta\to \infty}z=
\frac{[\alpha^{-1}-n_\alpha][1+n_\alpha]}{1+n_\alpha-\alpha^{-1}}
\label{eq:z_found}
\ee
This result is always positive and finite, so our ansatz indeed solves
the saddle-point equation. We can also calculate the zero
temperature limit of the fraction $c_n$ (\ref{eq:reducedstatsoln}) of sites with $n$
particles. Inserting $y=\beta \phi [n_\alpha+\frac{1}{2}]+\log(z)$
into (\ref{eq:reducedstatsoln}) gives
\be
c_\ell=\frac{{1\over \ell!}~z^{\ell}
e^{-\frac{1}{2}\beta\phi[\ell-\frac{1}{2}-n_\alpha]^2}}
{\sum_{m\geq 0}{1\over m!}~z^{m}
e^{-\frac{1}{2}\beta\phi[m-\frac{1}{2}-n_\alpha]^2}}
\ee
Upon taking the $\beta\to\infty$ limit, using (\ref{eq:z_found}), this reduces to
the expressions (\ref{eq:phil_1},\ref{eq:phil_2}) found earlier by direct variation of the $\{c_n\}$.

In contrast, the hydrophobic model  must maximize the variance in order to
minimize the energy (see (\ref{eq:energy_variance})), and for $\phi<0$ the system
will consequently compactify
(as we shall see) at any finite temperature, such that only $c_{n_c}$ and $c_0$ are
non-zero:
\begin{eqnarray}
\ell=0:   &&~
c_0=1\minus (\alpha n_c)^{-1}
\label{eq:phob_1}
\\
\ell=n_c: &&~
c_{n_c}=(\alpha n_c)^{-1}
\label{eq:phob_2}
\end{eqnarray}
with $c_\ell=0$ for all other $\ell$.
The variance then equals, in leading order as $n_c\to\infty$,
\be
\si^2=n_c/\alpha-1/\alpha^2
\label{eq:var_o_i}
\ee
and consequently diverges.
Apparently, for $n_c\to\infty$ the equilibrium state of the hydrophobic system
is one with just two types of site:
those with $n_c$ particles and those with no
particles (i.e. it is a maximally compact state), for any temperature.
We will now study the $n_c\to\infty$
behaviour of the saddle-point equation (\ref{eq:alpha}), for $\phi<0$ and arbitrary $\beta$.
We make the ansatz
\be
y=-1+\frac{1}{2}\beta \phi n_c-{1\over
n_c}\log(\al\sqrt{n_c})+\log(n_c)+x
\label{eq:sol_o_i}
\ee
where $\lim_{n_c\to\infty}x=0$.
This
converts (\ref{eq:alternative}) into
$\alpha^{-1}=\partial\log \tilde{Z}(x)/\partial x$, with,
upon application of Stirling's formula to the term $n=n_c$ in
$Z(y)$,
and modulo contributions to the exponents which vanish as
$n_c\to\infty$:
\begin{eqnarray}
\tilde{Z}(x)&=&1+\frac{e^{x n_c}}{\alpha
n_c\sqrt{2\pi}}e^{\order(\ldots)}
\nonumber \\
&&\hspace*{-5mm}
+\sum_{n=1}^{n_c-1}\frac{1}{n!} \left[e^{x-1
+\frac{1}{2}\beta |\phi|(n-n_c)-{1\over
n_c}\log(\al\sqrt{n_c})+\log(n_c)}\right]^n
\nonumber
\end{eqnarray}
One distinguishes two qualitatively different types of terms in
the above summation. Firstly, those with $n\ll n_c$  or with $n=\order(n_c)$
(provided $n\ll n_c$)
give a vanishing cumulative contribution to $\tilde{Z}(x)$ due to the
appearance of
$\frac{1}{2}\beta |\phi|(n-n_c)$ in the exponent.
Secondly, those terms with $n=n_c-k$, with finite $k>0$, each give
according to Stirling's formula
and modulo contributions to the exponents which vanish as
$n_c\to\infty$:
\bd
\frac{1}{n!} \left[e^{x-1
+\frac{1}{2}\beta |\phi|(n-n_c)-{1\over
n_c}\log(\al\sqrt{n_c})+\log(n_c)}\right]^n
\hspace*{10mm}
\ed
\bd
\hspace*{20mm}
=\frac{e^{x n_c}}{\alpha n_c\sqrt{2\pi}}
e^{-\frac{1}{2}n_c\beta|\phi|k
-k(x-1-\frac{1}{2}\beta|\phi|k)+
\order(\ldots)}
\ed
Hence also the second type of terms in $\tilde{Z}(x)$ are exponentially
damped as $n_c\to\infty$, relative to the two surviving terms $n=0$
and $n=n_c$, and hence
\begin{eqnarray}
\tilde{Z}(x)&=&
1+\frac{e^{xn_c}}{\alpha
n_c\sqrt{2\pi}}\left[1+\order(\ldots)\right]
\end{eqnarray}
Thus for $n_c\to\infty$
the saddle-point equation reduces to
\bd
\frac{1}{\alpha}=\lim_{n_c\to\infty}
\frac{n_c e^{xn_c}}
{\alpha n_c\sqrt{2\pi}+e^{xn_c}}
\ed
and hence $x=n_c^{-1}\log\sqrt{2\pi}+\ldots$ as $n_c\to \infty$,
confirming the correctness of our ansatz.
This gives for the fractions $c_n$ (\ref{eq:reducedstatsoln}) the
state (\ref{eq:phob_1},\ref{eq:phob_2})  which resulted earlier from direct maximisation of $\sigma^2$
by variation of $\{c_n\}$.

\section{The Transitions to Compact States}

We have observed in the simple single-species models of the previous
section  that in the hydrophilic case there are no phase
transitions for $n_c\to\infty$ (since here the entropic and energetic forces have
the same objective: a swollen state),
whereas in the hydrophobic case the system will for $n_c\to\infty$ be found in
a maximally compact state at any temperature (i.e. there is a phase transition at $T=\infty$).
Since the latter behaviour is strongly linked to the $n_c\to\infty$
limit, one should expect that proper phase transitions in the hydrophobic system
can be identified in terms of a suitably re-scaled $n_c$-dependent temperature (or, equivalently, a suitably re-scaled
interaction energy $\phi$). It
turns out the the proper scaling is obtained upon transforming
according to
\be
T= \tilde{T}|\phi| n_c^2/\log(n_c!),
\label{eq:be_red}
\ee
(where we have absorbed the interaction energy into the re-scaled temperature),
together with $\tilde{\beta}=1/\tilde{T}$.
For $n_c\to\infty$ we observed that the fraction  $c_{n_c}$ of sites with $n_c$ particles
vanished, even in the maximally compact state, where $c_{n_c}=(\alpha n_c)^{-1}$).
Hence  we switch in this section from inspecting the fraction $c_n$ (\ref{eq:reducedstatsoln}) of sites
with a given number $n$ of particles, to an alternative
representation of the macroscopic system state in terms of the fraction of particles
$f_n=\alpha n c_n$ which find themselves at a site filled with $n$
particles (in the maximally compact state one has $f_n=0$ for all $n<n_c$
and $f_{n_c}=1$). Note that $f_0=0$, by definition. We introduce the short-hand
\be
h(n,y)=\frac{\tilde{\beta} n^2\log(n_c!)}{ 2n_c^2}-\log(n!)+ny
\label{eq:define_h}
\ee
The saddle-point equation (\ref{eq:alpha}), from which to solve $y$, can now be written as
\be
\sum_{n=1}^{n_c}f_n=1,~~~~~
f_n=\frac{\alpha~n~e^{h(n,y)}}{\sum_{m=0}^{n_c}e^{h(m,y)}}.
\label{eq:nasty_saddle}
\ee
Equation (\ref{eq:nasty_saddle}) looks deceivingly simple.
However, it turns out that one cannot simply transform the
previous solution $y$ (\ref{eq:sol_o_i}) of the $\beta=\order(n_c^0)$ regime
by writing $\beta$ in terms
of the re-scaled $\tilde{\beta}$, due to the non-commuting of the
various operations involved.
Numerical solution of (\ref{eq:nasty_saddle}) indicates that for $n_c\to\infty$ we can
neglect all those terms $n$ in the summations with $1\ll n<n_c$,
relative to $n=n_c$, and that terms with $n\sim\order(1)$ are generally of
the same order of magnitude as that of $n=1$.
One can show that this property can be derived, from (\ref{eq:nasty_saddle}), as an immediate
consequence of assuming that the leading orders of $f_1$ and $f_{n_c}$ scale in the same way with
$n_c$ as $n_c\to\infty$. The latter, in turn,
is equivalent to saying $y=(1-\frac{1}{2}\tilde{\beta})\log n_c+u$
with $u$ finite for $n_c\to\infty$, exactly as in (\ref{eq:sol_o_i}) (after carrying out the
transformation from $\beta$ to $\tilde{\beta}$).
As a result, and in contrast to the scaling regime $\beta=\order(n_c^0)$,
even for $n_c\to\infty$  equation (\ref{eq:nasty_saddle}) remains
transcendental,
 and it generally resists analytical
solution.

A systematic numerical analysis of (\ref{eq:nasty_saddle}) reveals
the following picture. At high temperatures, $\tilde{\beta}=0$,
the system is in a swollen phase, which is defined by the condition
 $\lim_{n_c\to\infty}f_{n_c}=0$ (the fraction of monomers found at a maximally filled site is zero).
Upon increasing $\tilde{\beta}$ one then finds a phase transition
at $\tilde{\beta}^c_1(\al)$ to a partially compact state, where $0<\lim_{n_c\to\infty}f_{n_c}<1$.
A further increase of $\tilde{\beta}$ leads to a growth of the fraction of monomers in
maximally filled sites, until at a second transition temperature $\tilde{\beta}^c_2(\al)$
the system enters the fully compact phase, where
$\lim_{n_c\to\infty}f_{n_c}=1$.
For the special case $\tilde{\beta}=2$ it turns out that (\ref{eq:nasty_saddle})
can be solved analytically: for $\alpha>1$ one finds in leading order $y=-\log\alpha$ and $\lim_{n_c\to\infty}f_{n_c}=0$,
for $\alpha<1$ one has $y=\order(n_c^{-1}\log n_c)$ and $0<\lim_{n_c\to\infty}f_{n_c}<1$,
which confirms the picture obtained by numerical solution (see
below).

\begin{figure}[t]
\vspace*{12mm}\hspace*{0mm}
\setlength{\unitlength}{0.63mm}
\begin{picture}(0,90)
\put(-15,  0){\epsfxsize=140\unitlength\epsfbox{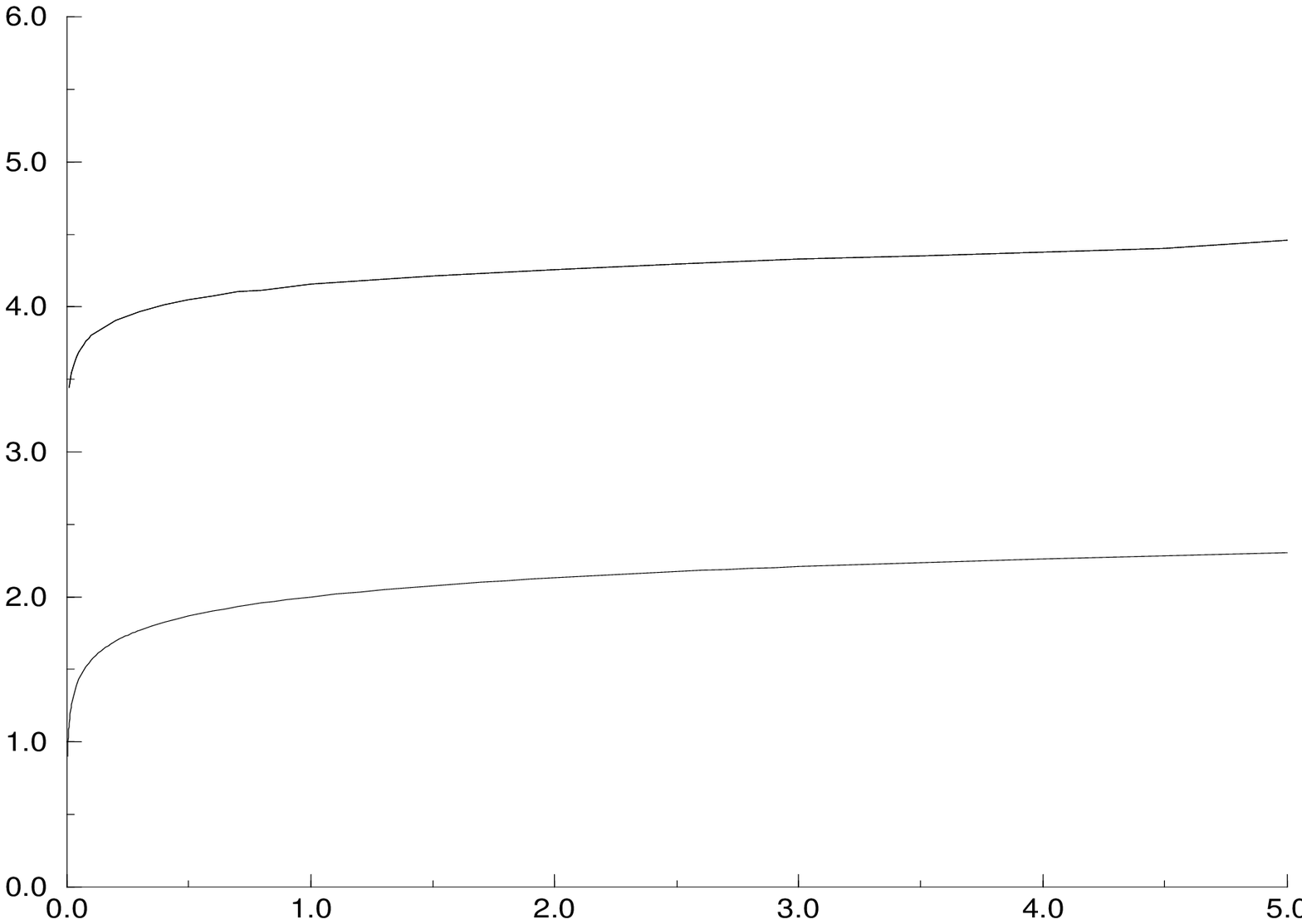}}
\put( 73, 8){{$\alpha           $}}
\put(  -2, 62){{$\tilde{\beta}          $}}
\put(120, 87){{$\tilde{\beta}^c_2(\al)$}}
\put(120, 57){{$\tilde{\beta}^c_1(\al)$}}
\put( 25, 90){{${\rm\bf\fns compact}~    (      f_{n_c}\!=\!1)$}}
\put( 25, 60){{${\rm\bf\fns partially~compact}~(0\!<\!f_{n_c}\!<\!1)$}}
\put( 25, 30){{${\rm\bf\fns swollen}~    (      f_{n_c}\!=\!0)$}}
\end{picture}
\caption{Phase diagram of the hydrophobic
model (with a single monomer species), as obtained by solving the saddle-point equation (\ref{eq:nasty_saddle})
numerically (with $n_c=2.10^5$). The re-scaled inverse temperature $\tilde{\beta}$ is defined in (\ref{eq:be_red}).
}
\label{fig:phase}
\end{figure}

The resulting phase diagram, obtained by extracting the two transition
temperatures $\tilde{\beta}^c_{1,2}(\al)$ from a numerical solution of
(\ref{eq:nasty_saddle}), is drawn in figure (\ref{fig:phase}), and is confirmed
qualitatively by numerical simulations (see the next section).
Note that even for the relatively high value
$n_c=2\times~10^5$ as used in the numerical evaluation, the line $\tilde{\beta}^c_2(\al)$
can  only be qualitatively correct, since in order to probe truly the $n_c\to\infty$ regime
we need  $\log(n_c)\to\infty$, according to the saddle-point equation.
Note, furthermore, that in the limit $\al\to 0$ the system must
be fully compact, so that both $\tilde{\beta} ^c_1$ and $\tilde{\beta} ^c_2$ must tend to $0$.
Finally, for $\al\gg 1$ our numerical solution shows that
 both transition temperatures
grow as $\tilde{\beta}_{1,2}(\al)\sim\log(\al)$.

\section{Numerical Simulations for High $n_c$}

%
\begin{figure}[t]
\vspace*{12mm}\hspace*{0mm}
\setlength{\unitlength}{0.62mm}
\begin{picture}(0,100)
\put(-15, 10){\epsfxsize=140\unitlength
              \epsfbox{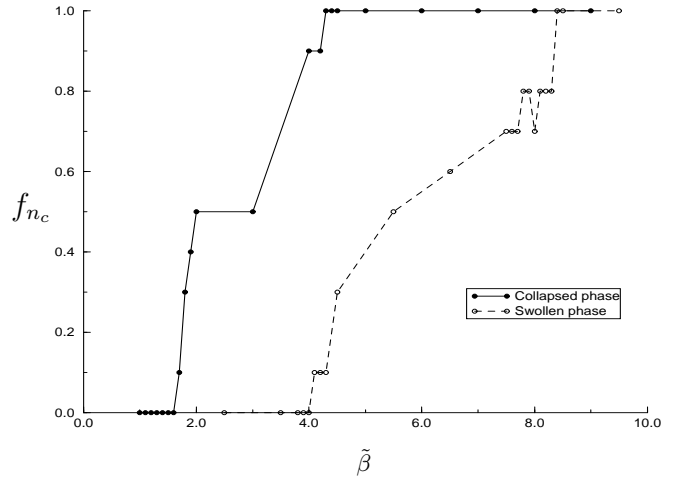}}
\put( 70, 16){$\tilde{\beta}$}
\put(  -4, 71){\large $f_{n_c}$}
\end{picture}
\vspace*{-5mm}
\caption{
Fraction $f_{n_c}$ of monomers in a maximally filled site,
measured in numerical simulations of the single-species hydrophobic
system ($\phi<0$), as a function of the re-scaled inverse temperature.
Simulation parameters: $N=R=1000$ (so $\alpha=1$), $n_c=100$, and
relaxation time $t=10^6$ iterations/monomer.
Initial conditions: either the compact state (solid), or the
swollen state (dashed).
}
\label{fig:hyst_t=million}
\end{figure}
We have carried out numerical simulations to verify the
theoretical equilibrium results of the last section, and the kinetics of the approach to equilibrium
in the $n_c\to\infty$ regime;
since phase transitions
only occur for $\phi<0$ (the hydrophobic system) we have restricted our simulations
accordingly.
Due to often excessive equilibration times observed for large $n_c$ (see below),
it has been necessary to limit our experiments to $n_c=100,~N=1000$. Hence, strong
finite size effects could be anticipated, since the theory requires $N\to
\infty$, $n_c\to\infty$ and $n_c/N\to 0$. Nevertheless, the numerical simulations were found to
agree remarkably
well with the theory.

Figure \ref{fig:hyst_t=million} shows the fraction $f_{n_c}$ of monomers in
a maximally filled site, as a function of the re-scaled inverse temperature
$\tilde{\beta}$, after $10^6$ updates per monomer. The solid curve
shows the observed value following initialisation of the system in a maximally compact (collapsed) state;
the dashed curve shows the observed $f_{n_c}$ following initialisation in
the fully swollen state.
The hysterisis-type graph which results from these measurements
(the dependence on intialisation) reflects
extremely slow equilibration of the system towards more compact states, due to entropic
barriers (similar to those in e.g. \cite{ritort}).
The full curve (which describes a process of de-compactification) does not suffer
 from this problem,
and therefore gives the best experimental estimate for the true locations of the transitions.
This latter curve predicts (for $\al=1.0$) that $\tilde{\beta}_1^c(1)\simeq 1.7$ and $\tilde{\beta}_2^c(1)\simeq 4.1$,
to be compared to the theoretical predictions $\tilde{\beta}_1^c(1)\simeq 2.0$ and $\tilde{\beta}_2^c\simeq 4.2$
(see figure \ref{fig:phase}).
Given the anticipated finite size effects, these values are remarkably close.
The non-monotonicity of some of the experimental curves are surely finite $n_c$
effects; for finite $n_c$ a maximally filled site can indeed temporarily
loose one or more monomers, whereas the probability for this to happen vanishes
for $n_c\to\infty$.

\begin{figure}[t]
\vspace*{18mm}\hspace*{0mm}
\setlength{\unitlength}{0.62mm}
\begin{picture}(0,70)
\put(-15,-10){\epsfxsize=140\unitlength\epsfbox{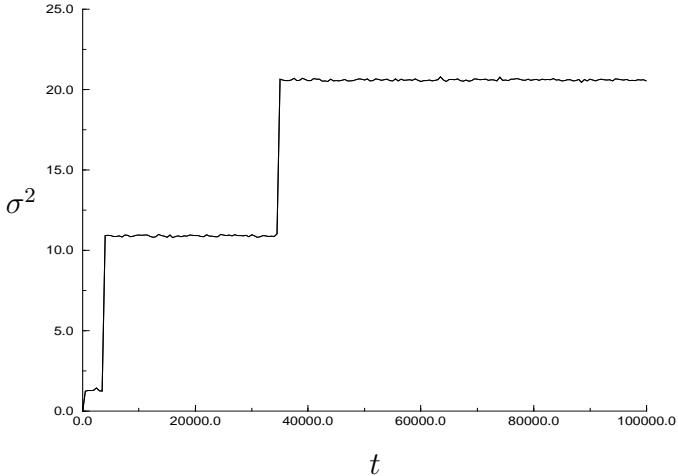}}
\put( 73,  -5){\large $t$}
\put(  -5, 51){\large $\sigma^2$}
\end{picture}
\vspace*{7mm}
\caption{Variance $\sigma^2=\bra n^2\ket-\bra n\ket^2$ (proportional to the energy)
as a function of time, measured in numerical simulations of a hydrophobic system
as in the previous figure, at $\tilde{\beta}=3.8$, following fully swollen initial conditions.}
\label{fig:plateau}
\end{figure}

\begin{figure}[t]
\vspace*{46mm}\hspace*{13mm}
\setlength{\unitlength}{0.88mm}
\begin{picture}(0,165)
\put(-15,150){\epsfxsize=87\unitlength\epsfbox{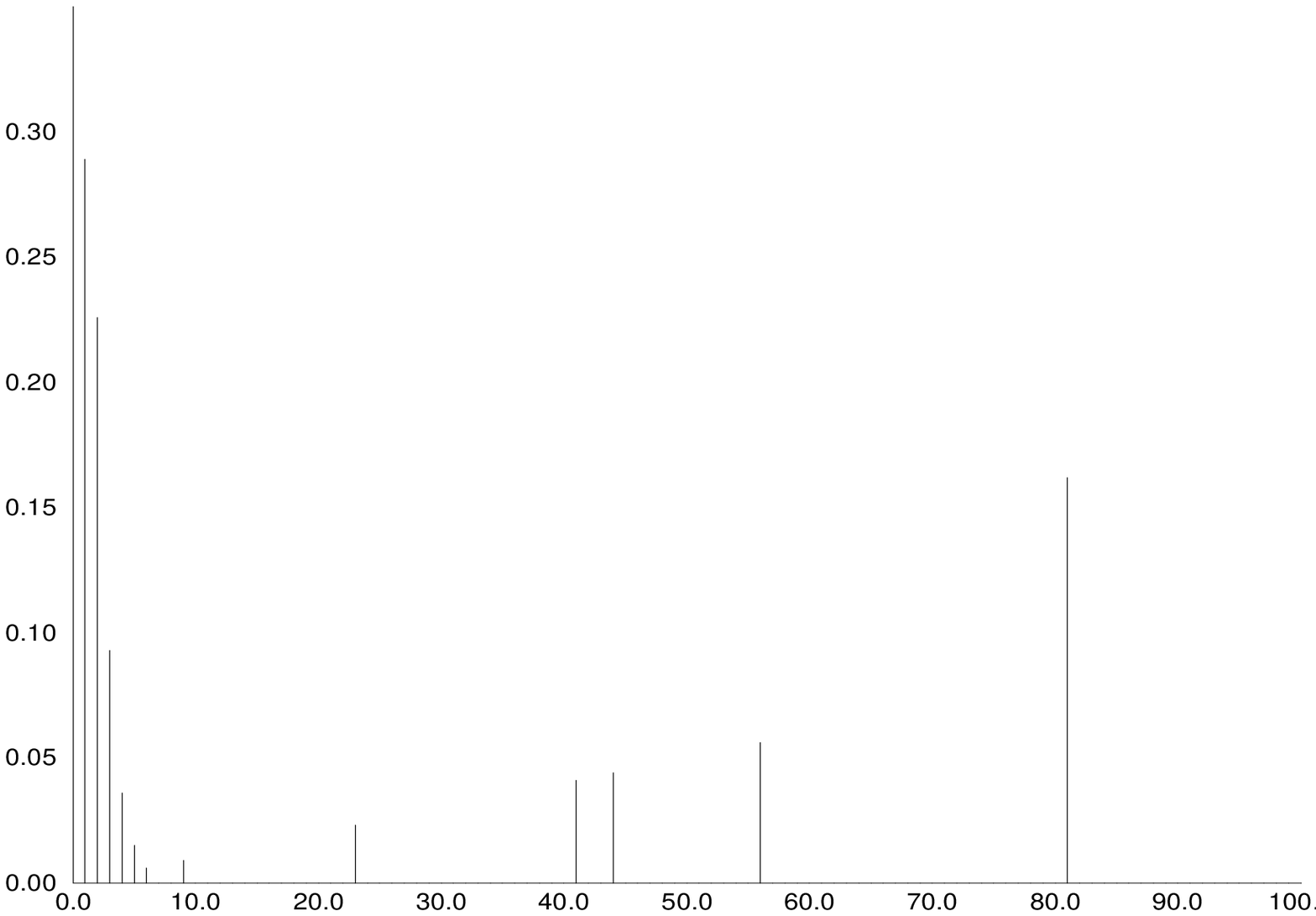}}
\put(-15, 80){\epsfxsize=87\unitlength\epsfbox{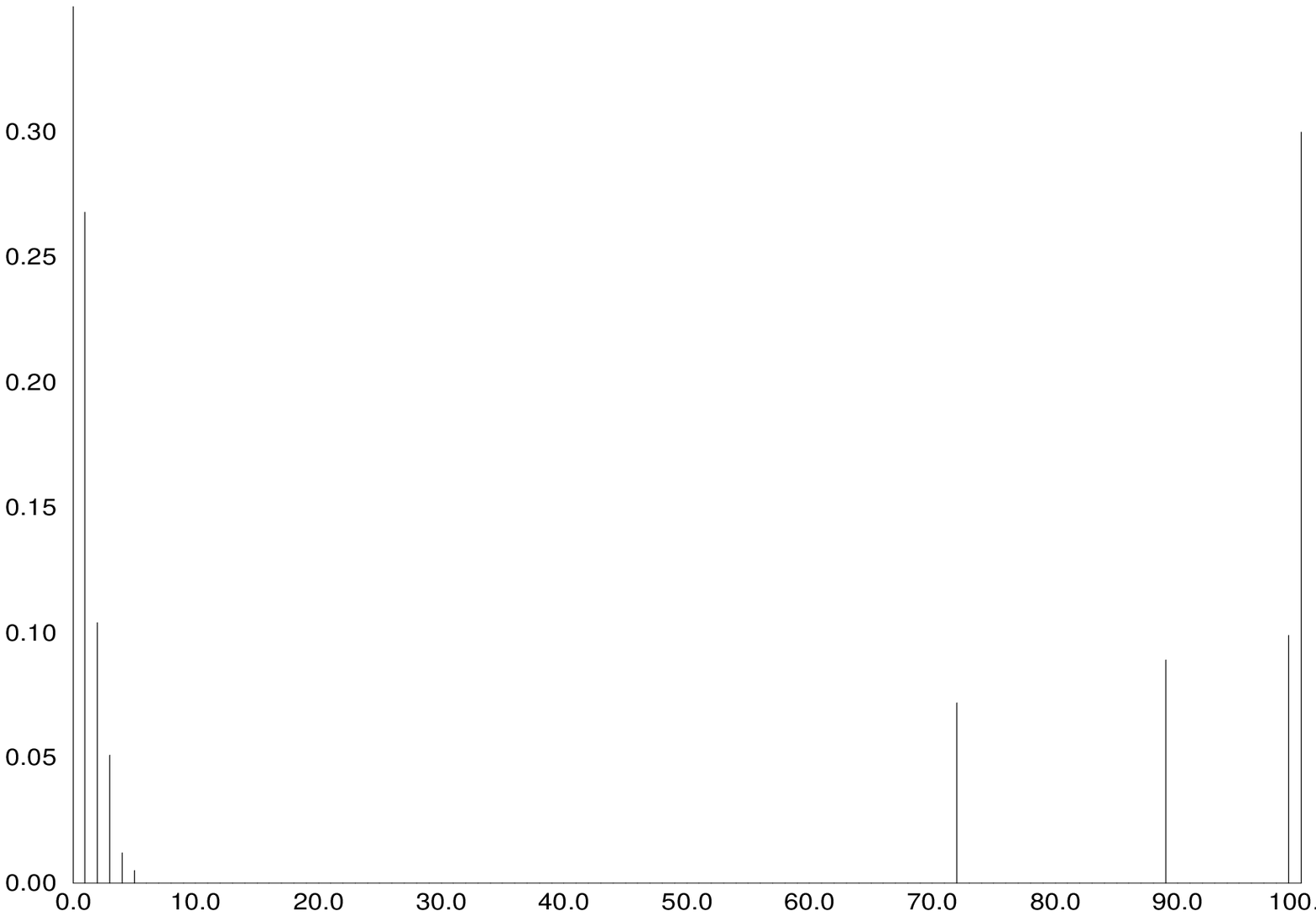}}
\put(-15, 10){\epsfxsize=87\unitlength\epsfbox{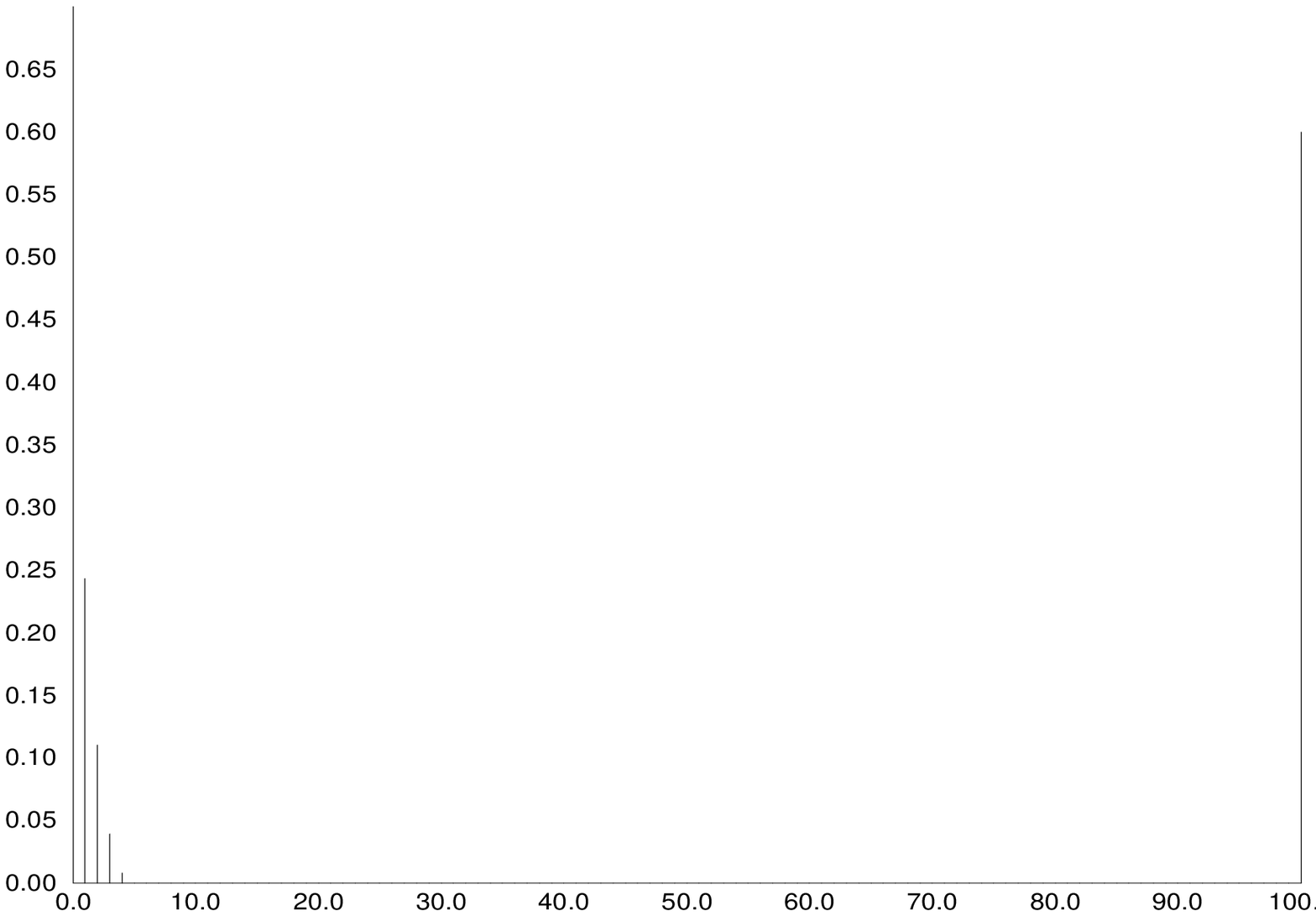}}
\put( 36,153){\large $n  $}
\put( 36, 83){\large $n  $}
\put( 36, 13){\large $n  $}
\put( -17,190){\large $f_n(t_1)$}
\put( -17,120){\large $f_n(t_2)$}
\put( -17, 50){\large $f_n(t_3)$}
\end{picture}
\vspace*{-7mm}
\caption{The occupation fractions $f_n$ as observed in a single numerical simulation
at times $t_1\!=\!10^2,~t_2\!=\!2~10^2,~t_3\!=\!10^4$, for a hydrophobic system with
$N=R=1000,~n_c=100$, and $\tilde{\beta}=8.0$. It illustrates the
formation, within a relatively small number of iteration steps, of a maximally filled site. }
\label{fig:3graphs}
\end{figure}

\begin{figure}[t]
\vspace*{12mm}\hspace*{1mm}
\setlength{\unitlength}{0.62mm}
\begin{picture}(0,100)
\put( -15,  10){\epsfxsize=140\unitlength\epsfbox{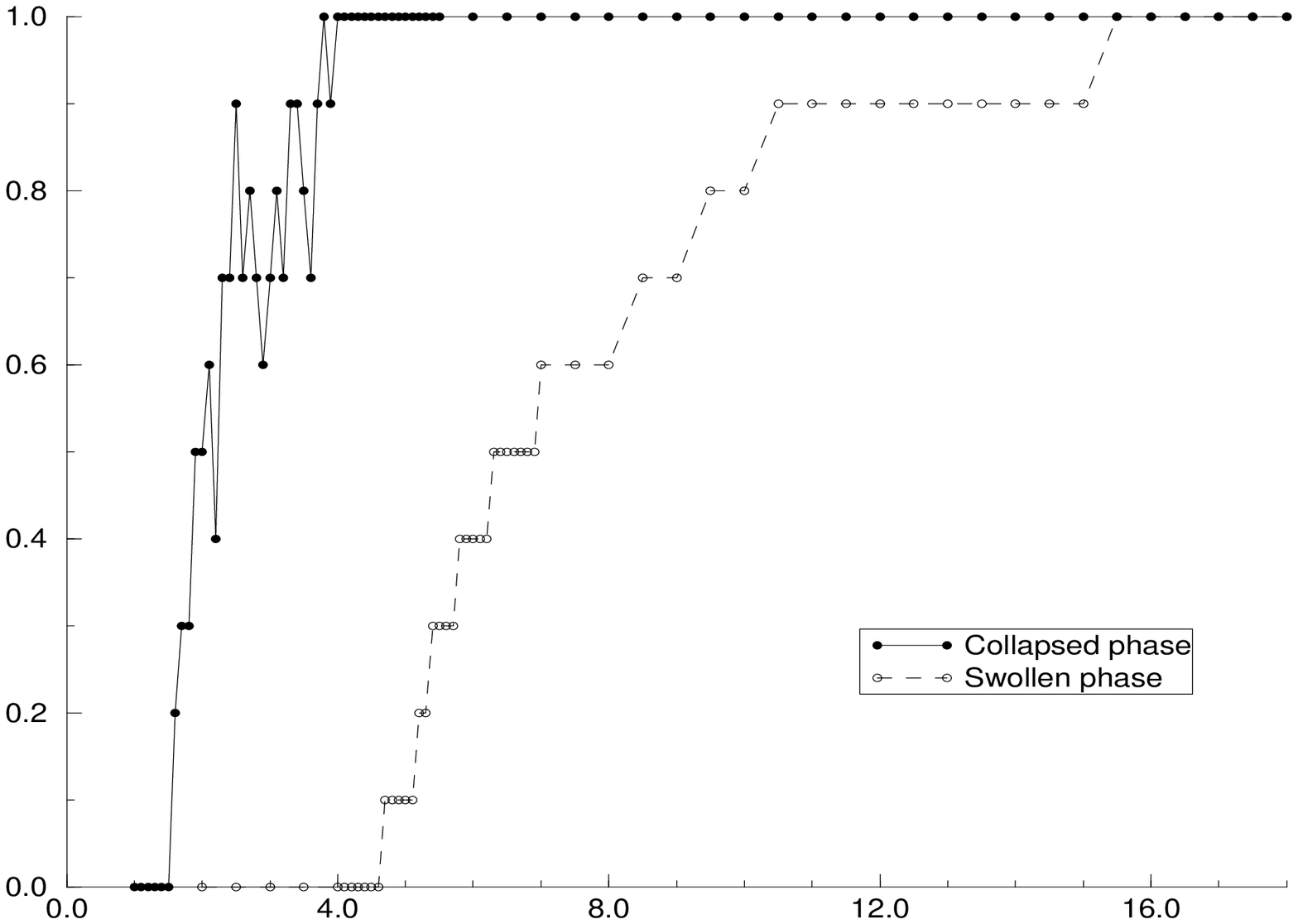}}
\put( -15,-100){\epsfxsize=140\unitlength\epsfbox{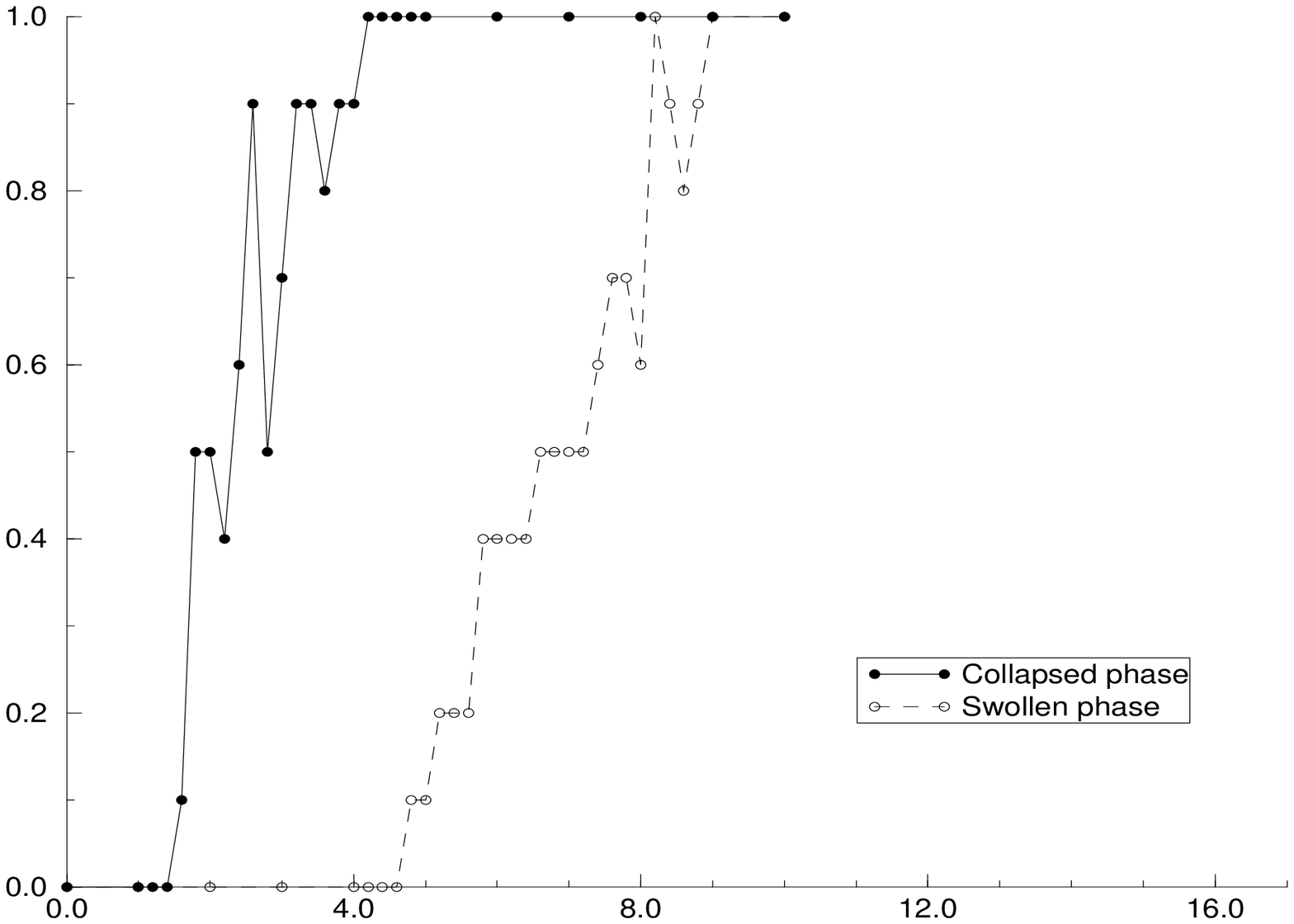}}
\put(  67, 15){$\tilde{\beta}$}
\put(  -5, 71){\large $f_{n_c}$}
\put(  67, -95){$\tilde{\beta}$}
\put(  -5, -39 ){\large $f_{n_c}$}
\end{picture}
\vspace*{65mm}
\caption{Occupation fractions $f_{n_c}$ versus the re-scaled inverse temperature
$\tilde{\beta}$, as measured at $t=1~10^5$ in simulations of a hydrophobic system with $N=R=1000,~n_c=100$.
Upper graph: the system is not re-initialized when starting a new simulation at increased $\tilde{\beta}$.
Lower graph: for each  $\tilde{\beta}$ the system is re-initialized
to a maximally compact or swollen state.}
\label{fig:hyst}
\end{figure}

We now turn to the entropy barriers which cause the slow equilibration
towards compact states.
At sufficiently low temperatures, starting from a swollen (i.e. high temperature) state,
monomers initially hop around until at some site a critical (temperature dependent)
monomer number is obtained. After this has happened, this site fills up very rapidly, since
its monomers can no longer escape (for energetic reasons).
All monomers in fully filled sites are thus more or less immobile, and the remaining (mobile)
monomers find themselves at a reduced effective density
$\rho_{\rm eff}=(1-f_{n_c})/(\alpha-c_{n_c})$. Hence, as $f_{n_c}$ increases,
for the mobile monomers it will become increasingly harder to
`find' other monomers to form a new site with a critical monomer number. This entropic effect is
illustrated by the simulation shown in in  figure \ref{fig:plateau}. It is characterised by
sudden jumps in the  variance, corresponding to a site reaching the critical monomer
number and filling up rapidly, with increasingly long plateaus in between,
where the remainder of mobile monomers hop around at a reduced effective density.
The actual process of the rapid filling of those sites which a some stage reach the critical monomer number,
is illustrated in figure \ref{fig:3graphs}.
It shows, as snapshots of numerical simulations
at three intermediate (relatively close) times, the distribution $\{f_n\}$
for a hydrophobic system as in the previous experiments which has been
quenched from a maximally swollen state to a lower temperature, well within the fully
compact phase. Since the equilibrium state in that phase corresponds to $f_{n_c}=1$,
even at the largest observation time shown, $t_3=10^4$, the system clearly
has not yet equilibrated.

Another interesting aspect of the equilibration dynamics is the fact that
compactification is found to take longer for a system which is not re-initialised at each
new relaxation temperature, compared to a system that is. This is illustrated in figure
\ref{fig:hyst}.
Apparently, in a quench (re-initialisation to a swollen state), several sites
reach the critical monomer density simultaneously, while the effective density
is still reasonably high. Note that at short times one can indeed observe
several peaks in figure \ref{fig:3graphs}. On the other hand, in a
system which is not
re-initialised, the effective monomer density is already reduced when the temperature
is changed and a new relaxation commences.
As a result, the entropic barrier to be overcome before a critical monomer density at any site
is achieved, is larger in the system which is not re-initialised than in the re-initialised system,
and hysteresis effects are therefore stronger. Non-monotonicity is, again, a finite $n_c$ effect.

\section{Discussion}

In this paper we have studied the dynamics and the compactification phase transitions
of a class of simple lattice gas models
for a random heteropolymer in infinite dimensions, as introduced in
\cite{jort}.
Within this class we have restricted ourselves to models with site
disorder, in particular to the Random Hydrophobic Model (RHM).
For finite $n_c$ (the maximum number allowed at a given site), and
a finite number of monomer species,
we have been able to solve the dynamics exactly, by deriving exact closed macroscopic
evolution equations for a suitable set of dynamic order
parameters.
The solution of the (non-linear) macroscopic equations, describing the approach to
equilibrium, is shown
to be in excellent agreement with numerical simulations.

In order to study the physics of the model also for $n_c\to\infty$ (a limit
which could not easily be taken in the equilibrium solution proposed
in \cite{jort}) we have presented an alternative (but equivalent) equilibrium
solution. In the latter approach the limit $n_c\to\infty$ poses no
problems (although the method would not apply to models with bond
disorder, in contrast to that of \cite{jort}).
Working out our equations for  models which just a single monomer
type shows that for $n_c\to\infty$ a hydrophilic model will not
undergo phase transitions at any temperature (as expected), but that the hydrophobic
system will be in a collapsed state, at any finite temperature.
In the latter case the transition(s) apparently occur at
$T=\infty$, as a consequence of having taken the $n_c\to\infty$
limit. In terms of a suitably re-scaled inverse temperature $\tilde{\beta}$, however,
two transitions are found to occur,
at $\tilde{\beta}_1^c$ (from a swollen to a partially compact state) and at
$\tilde{\beta}_2^c$ (from a partially compact to a fully compact state).
Both the existence and location of these two transitions
are confirmed by numerical simulations, within the experimental
limits imposed by having finite $N$ and finite $n_c$ (the theory requires $N\to\infty$, $n_c\to\infty$
with $n_c/N\to 0$). Especially when following a swollen (i.e. high temperature)
initial state, equilibration is
observed to take place via an extremely slow plateau-like dynamics,
which is caused by entropic effects,
particularly at the critical points and in the partially compact phase.

A natural extension of the work reported in this paper would be to solve
the dynamics of the model for the case of bond
disorder. In statics such models have been solved using replica theory, see
\cite{jort,GOT,GOP,Ob,GLO,GO_C,Sf}, and it is to be anticipated that the method of path integrals
\cite{pathint} would have to be used  for solving the dynamics,
as in e.g.
\cite{TAB}.

Infinite dimensional models are obviously less realistic
than finite dimensional ones in terms of the spatial properties of
polymer folding processes. However, our present study shows that,
due to the absence of any need for approximations
(in contrast to their finite dimensional counterparts),
they do allow for a fairly complete and fully exact quantitative understanding of the
energetic and entropic interplay between hydrophobic and
hydrophilic effects, including both statics and dynamics.


\end{document}